%% file: Article.tex
\documentclass[aps,prd,preprint,showpacs,superscriptaddress]{revtex4-2}
\usepackage{amsfonts}
\usepackage{amsmath}
\usepackage{amssymb}
\usepackage{dsfont}
\usepackage{hyperref}
\usepackage{graphicx}
\usepackage{mathtools}
\usepackage[usenames,dvipsnames]{xcolor}
\usepackage{times}
\usepackage{braket}
\usepackage{booktabs}
\usepackage{multirow}
\usepackage{tikz}
\usepackage{placeins}
\usepackage[font=normal, justification=Justified]{caption}

\newcommand{\orcid}[1]{\href{https://orcid.org/#1}
	{\includegraphics[width=7pt]{orcid.png}}}

\hypersetup{    colorlinks=true,linkcolor=blue,citecolor=blue,filecolor=blue,urlcolor=blue,breaklinks=true}
\RequirePackage{color}

\begin{document}

\title{Remarks on the Study of Electronic Properties and Photoionization Process in Rotating 2D Quantum Rings}

\author{Carlos Magno O. Pereira}
\email{cmop302@gmail.com}
\affiliation{Departamento de F\'{\i}sica, Universidade Federal do Maranh\~{a}o, 65085-580 S\~{a}o Lu\'{\i}s, Maranh\~{a}o, Brazil}

\author{Frankbelson dos S. Azevedo}
\email{frfisico@gmail.com}
\affiliation{Departamento de F\'{\i}sica, Universidade Federal do Maranh\~{a}o, 65085-580 S\~{a}o Lu\'{\i}s, Maranh\~{a}o, Brazil}

\author{Edilberto O. Silva}
\email{edilberto.silva@ufma.br}
\affiliation{Departamento de F\'{\i}sica, Universidade Federal do Maranh\~{a}o, 65085-580 S\~{a}o Lu\'{\i}s, Maranh\~{a}o, Brazil}

\date{\today}

\begin{abstract}
Electronic and optical properties of a mesoscopic heterostructure of a two-dimensional quantum ring composed of Gallium Arsenide (GaAs) semiconductors are investigated. Using the confinement potential proposed by Tan-Inkson to describe the system under analysis, we conducted a numerical study of the photoionization cross section for a 2D quantum ring with and without rotation effects. The interior of the quantum ring is traversed by an AB flux. Our research aims to investigate how this mesoscopic structure's electronic and optical properties respond to variations in the following parameters: average radius, AB flux, angular velocity, and incident photon energy. Under these conditions, we establish that optical transitions occur from the ground state to the next excited state in the conduction sub-band, following a specific selection rule. One of the fundamental objectives of this study is to analyze how these rules can influence the general properties of two-dimensional quantum rings. To clarify the influence of rotation on the photoionization process within the system, we offer findings that illuminate the effects of the pertinent physical parameters within the described model. We emphasize that, although this is a review, it provides critical commentary, analysis, and new perspectives on existing research. Some results presented in this paper can be compared with those in the literature; however, new physical parameters and quantum ring configurations are used.
\end{abstract}
\maketitle

\section{Introduction}\label{sec1}

Due to the advancement and rapid improvement of nanostructure technology, electronic and optical properties related to low-dimensional semiconductor systems have attracted great interest from researchers in the field of mesoscopic physics, an area dedicated to the investigation of quantum systems on an intermediate scale, characterized by dimensions larger than the atomic scale but still below the limits defining macroscopic objects \cite{ferry2015transport,vladimirphysics}. Such systems have dimensions on the order of nanometers, and they are the subject of extensive investigation not only due to their scientific relevance but also because of their wide technological applicability in optoelectronic devices, generating significant interest in both theoretical \cite{PhysRevB.50.8460} and experimental studies \cite{PhysRevLett.84.2223}.

As a result, various electronic and optical properties have been explored in the context of quantum confinement within mesoscopic systems, such as Quantum Wells (QWs) \cite{bahar2021optical,chen2008nonlinear,enders2004electronic}, Quantum Dots (QDs) \cite{PE.2023.147.115617,AdP.2019.531.1900254,candemir2023linear,chubrei2021effect,pal2019impurity,perez2023intense}, Quantum Rings (QRs) \cite{FBS.2022.63.64,AdP.2022.535.2200371,PE.2024.158.115898,PE.2021.132.114760,Mingge2013,AoP.2023.459.169547}, among other mesoscopic structures \cite{FBS.2022.63.58,poole2003introduction}. These structures are generally composed of two or more layers of different types of semiconductors juxtaposed, known as heterostructures, where there is a confinement of charge carriers \cite{PRB.1997.55.15688,PRB.1998.57.6584,JPCM.2022.34.105302}. Recently, M. G. Bawendi, L. Brus, and A. Ekimov were awarded the Nobel Prize in Chemistry for their discovery and synthesis of QDs, characterized by quantum confinement in all three spatial dimensions. This confinement leads to the complete localization of charge carriers, resembling the behavior of atoms. Consequently, this arrangement results in the manifestation of discrete energy levels \cite{pal2019impurity,HEYN2021}, which is why QDs are commonly referred to as artificial atoms in the literature \cite{chakraborty1999quantum,PhysRevLett.65.108,BOSE1998238,CAKIR20122659}.  

QDs and QRs are manufactured from semiconductor heterostructures and are crucial for advancements in nanotechnology, particularly in chip manufacturing \cite{biasiol2011compositional}. They are indispensable components in cell phones, computers, and even cars. These materials are also used in color-illuminating technologies like televisions and Light Emitting Diode (LED) lamps and in various electronic and optical devices. Additionally, they play a significant role in medical applications, such as guiding surgeries for the removal of tumor tissues \cite{doi:10.1021/acsanm.0c01386,doi:10.1021/acs.jpclett.7b00671,ameta2022quantum}.

From a physics point of view, semiconductors are a group of materials with intermediate electrical conductivity between metals and insulators, described by band theory \cite{madelung2012semiconductors,yu2010fundamentals}. The conductivity of semiconductors can be adjusted, enabling them to either allow the passage of electric current, acting like metals, or to prevent it, functioning as insulators. Semiconductors are classified into two general categories: elemental semiconductor materials, found in group IV of the periodic table, and compound semiconductor materials, most of which are formed from special combinations of elements from groups III and V \cite{neamen2002semiconductor}.

\begin{figure}[tbh]
    \centering
    \includegraphics[scale=1.2]{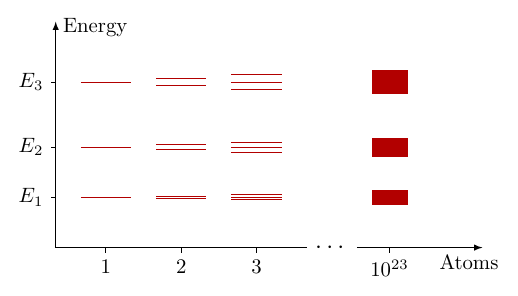}\\
     (a) Energy levels of atoms in atomic interdependence.
       \includegraphics[scale=1.2]{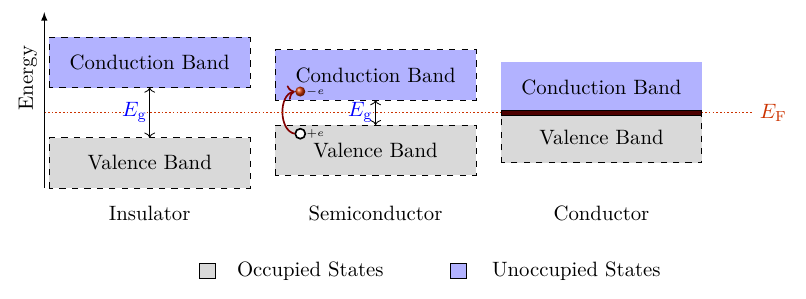}\\
        (b) Gap between the valence and conduction bands of semiconductor, insulator, and conductor. \cite{Book.Heinzel.2008,Book.Kittel.2007}.
    \caption{Schematic representation of the formation of energy bands in materials.}
    \label{fig:comparison}
\end{figure}

As in solids, atoms in semiconductors are very close to each other, with interatomic separations approximately equal to the size of the atoms, leading to significant interactions. In an isolated atom, electrons have specific allowed energy levels they can occupy. However, as the number of atoms increases, the outer orbitals overlap and interact strongly with each other. This interaction broadens the discrete energy levels of isolated atoms into continuous bands, as schematically represented in Figure \ref{fig:comparison}(a). The last fully occupied electronic band is called the valence band, while the next higher band is known as the conduction band, which can be either completely unoccupied or partially filled. The energy difference between the minimum of the conduction band and the maximum of the valence band is called the band gap, denoted as $E_{\mathrm{g}}$, where no allowed energy states exist. Figure \ref{fig:comparison}(b) illustrates a generic diagram of the energy levels, depicting the valence and conduction bands of insulators, semiconductors, and conductors \cite{Book.Kittel.2007}.

Insulators are materials that have the last band completely filled. Applying an external electric field does not alter the electrons' total momentum (zero) since all states are occupied. Therefore, no electric current occurs. The energy level above which there are no occupied states at a temperature of $T=0~\mathrm{K}$ is designated as the Fermi level ($E_\mathrm{F}$) \cite{Book.Kittel.2007} (see Fig. \ref{fig:comparison}(b)).

For semiconductors at absolute zero temperature ($T=0 \mathrm{~K}$), electrons completely fill all energy bands, starting from the lowest to the valence band. As the temperature increases, a certain number of electrons can be excited and gain sufficient energy to jump from the valence band to the conduction band (see Fig. \ref{fig:comparison}(b)), contributing to the material's conductivity. In this process, electrons leave behind empty states, called holes, which behave like positively charged particles, moving in the opposite direction to electrons. This type of conduction, where electrons and holes carry current, is characteristic of semiconductors \cite{Book.Kittel.2007}.

Conductive materials, also called metals, exhibit high electrical and thermal conductivity, maintaining the ability to conduct electric current even at absolute temperature ($T=0~\mathrm{K}$). Figure \ref{fig:comparison}(b) illustrates a schematic representation of metals, showing the condition where the last valence band is partially filled. Charge carriers move practically freely when an electric field is applied to these materials. The absence of the forbidden region can explain the remarkable conductivity of metals. These characteristics of metals are strategically exploited and applied in the conduction of electric current, especially in the electricity sector's infrastructure \cite{NDTI.2015.75.33,gonen1986electric}.

A recent surge of interest in QRs has emerged due to their unique topological properties and potential applications in various cutting-edge technologies \cite{CTP.2024.76.105701,AdP.2023.535.2200371}. QRs, unlike QDs, present a toroidal structure, enabling the investigation of phenomena such as magnetization, persistent currents, and the Aharonov-Bohm (AB) effect, where the quantum states of the system can be influenced by a magnetic flux threading the ring without the presence of a classical force \cite{aharonov1959significance}. This phenomenon has motivated extensive research in both theoretical (in the studies of optical properties, such as refractive index changes, absorption coefficients, photoionization process, and also in thermodynamic properties) \cite{SM.2013.58.94,PRB.2011.84.235103,PRB.2013.87.035314,PRB.1995.52.1932,ASCT.2021.30.62,EPJP.2019.134.459,Nature.2001.413.822,PRB.2005.71.033309,PLA.2013.377.903,PRB.2018.98.205408,ECMP.2024.SE.415,PLA.2012.376.1269,EPJPlus.2014.129.147,LIANG20115818,Aghoutane2018,EPJD.2023.77.143,JTAP.2023.17.2,PB.2024.673.415438,MP.2024.0.2338400,MP.2022.120.2046295,PE.2021.131.114710,JLTP.2021.202.83} and experimental \cite{NL.2018.18.6188,PRB.2017.95.205426,NTec.2004.15.s126,OQE.2022.54.463,PRB.2011.83.115448} contexts, as it offers a rich platform for studying the interplay between quantum confinement, geometry, and magnetic fields. 

Furthermore, the introduction of rotation in these systems enhances their physical properties, and this is one of the main topics that will be explored in this work. A rotating quantum ring, particularly in a 2D setting, leads to new physics due to the coupling between rotational motion and magnetic fields. In this scenario, the Coriolis and centrifugal forces modify the effective potential experienced by the charge carriers, affecting the energy levels \cite{EPL.2015.110.27003,RSPA.2015.472.0858,RP.2015.5.55,EPJP.2019.134.546,CTP.2024.76.065504,IJTP.2015.54.2119} and optical properties of the system \cite{PhysRevB.91.035308,PhysRevB.96.165303}. 
Rotating QRs also holds promise for developing spintronic devices, where manipulating spin degrees of freedom is crucial for information processing and storage. The ability to control the spin of charge carriers in mesoscopic systems through external fields, including magnetic and rotational fields, makes QRs ideal candidates for quantum bit (qubit) implementations in quantum computing. Based on persistent currents in QRs, topologically protected qubits provide a robust mechanism for quantum information storage and manipulation \cite{Alicea2011}.

In addition to these technological applications, rotating QRs offer a promising platform for exploring fundamental quantum mechanical effects. The interplay between magnetic fields and rotational motion in these systems can lead to novel quantum phenomena, such as the appearance of Berry phase effects, which play a central role in understanding topological materials and their electronic properties. This makes QRs a valuable tool for investigating topological insulators \cite{PLA.2019.383.125865}, quantum Hall effects in curved geometries \cite{SPC.2022.5.029,PRB.2007.76.125430,AoP.2015.362.752}, and other exotic quantum states of matter \cite{sarkar2013exotic}.

Overall, the study of electronic properties and photoionization cross-sections in rotating 2D QRs opens new avenues in both applied and fundamental research. These systems exhibit a rich interplay between quantum confinement, magnetic fields, and rotation, leading to the manifestation of novel quantum phenomena with potential applications in optoelectronics, spintronics, and quantum computing.

In the following sections of this paper, we investigate how variations in physical parameters such as average radius, AB flux, angular velocity, and incident photon energy affect the electronic properties and photoionization processes. In Section \ref{sec2}, we conduct a detailed review of the Tan-Inkson model by exploring the formation of energy bands for physical parameter values that have not been previously evaluated in the literature. We also carefully evaluate the probability distribution and confinement potential for the QRs. In Section \ref{sec3}, we analyze the photoionization process by utilizing the energy states and wave functions obtained in the previous section. In Section \ref{sec4}, we introduce the effects of rotation on the Tan-Inkson model and assess their impact on the energy spectra and probability distributions. Finally, Section \ref{sec5} presents the summary and conclusions. 

\section{The Tan-Inkson model}\label{sec2}

Tan and Inkson introduced a theoretical model to investigate electron states in two-dimensional QRs, employing an exact solvable approach \cite{SST.1996.11.1635,PRB.1999.60.5626}. The theoretical model describes electron behavior in a confined two-dimensional ring structure, exploring the quantized properties arising from geometric constraints. The proposed model accounts for electron interactions within the two-dimensional ring, considering the effects of confinement on electron energy levels. The investigation delves into how properties such as ring radius and applied potentials influence the electronic states within the ring. The article derives exact solutions to the equations describing these electronic states, offering precise insights into electron behavior in this specific system. The obtained results have implications for the design and understanding of devices based on rings or similar nanostructured systems \cite{JPC.2023.7.045002,EPJP.2019.134.495,CPL.2022.806.140000,GRG.2019.51.120,PLA.2019.383.1110,PLA.2016.380.3847,PBCM.2015.459.36,EPJB.2012.85.354,PLA.2012.376.1269,PRB.2011.84.075332}. This work contributes to the development of more efficient and advanced electronic devices. Overall, this study sheds light on the quantized behaviors of electrons in nanostructured systems, providing valuable insights for applications in electronic devices \cite{vladimirphysics,PRB.2013.87.035314}.

In the model, the Hamiltonian that describes the quantum system considers an electron confined to a two-dimensional quantum ring in the effective mass approximation structure and the presence of a uniform static magnetic field $\boldsymbol{B}$ along the $\boldsymbol{z}$ direction, perpendicular to the ($x$-$y$) plane of the ring. Additionally, there is an infinite thin magnetic flux (AB flux) $\phi=l\phi_{0}$ (where $\phi_{0}=h/e$) crossing the center of the $2\mathrm{D}$ ring, with the parameter $l$ being a continuous quantity. The system's Hamiltonian is then written as follows
\begin{align}
	H & =\dfrac{1}{2\mu}\left(\boldsymbol{p}-e\boldsymbol{A}\right)^{2}+V(r),\label{eq:hamiltonianaQR2D1}
\end{align}
where $\mu$ and $e$ are the electron's effective mass and charge. The vector potential associated with the effective magnetic field is given by 
\begin{equation}
	\boldsymbol{A}=\frac{1}{2}Br\boldsymbol{\hat{\varphi}}+\frac{l\hbar}{er}\boldsymbol{\hat{\varphi}}.\label{pv}
\end{equation}
The confinement potential $V(r)$ includes an approximation for a parabolic and inverse square distance type potential given by \cite{SST.1996.11.1635,PRB.1999.60.5626}
\begin{equation}
	V(r) =\dfrac{a_{1}}{r^{2}}+a_{2}r^{2}-V_{0},\label{eq:v(r)inkson}
\end{equation}
where $V_{0}=2\sqrt{a_{1}a_{2}}$, and $a_{1}$ and $a_{2}$ are parameters that can be adjusted independently through appropriate choices. The freedom to choose these parameters leads us to access other mesoscopic models of interest.
Table \ref{tab1} shows the models we can study using the potential (\ref{eq:v(r)inkson}) by adjusting $a_{1}$ and $a_{2}$.

\begin{table}[htbp]
	\centering
	\caption{Structures modeled by the Tan-Inkson potential} \label{tab:exemplo_booktab}
	\begin{tabular}{|p{0.25\linewidth}p{0.7\linewidth}|}
		\hline
		\textbf{Quantum Dot} & In the limit $a_{1}\rightarrow0$, we obtain the potential
		$V(r) \rightarrow V_{QD}(r)=a_2 r^2.$
		\\
		\hline
		\textbf{Quantum Antidot} & In this case, for $a_{2}\rightarrow0$, we obtain 
		$V(r) \rightarrow V_A(r)=\frac{a_1}{r^2}.$ \\
		\hline
		\textbf{2D Infinite Wire} & This structure is obtained by making $r_0 \rightarrow \infty$.\\
		\hline
		\textbf{1D Quantum Ring} & Taking $a_2 \longrightarrow \infty$, we have $\omega \longrightarrow \infty$, by definition. \\
		\hline
	\end{tabular}\label{tab1}
\end{table}

\begin{figure}[tbh]
	\centering
    \includegraphics[height=5.5cm]{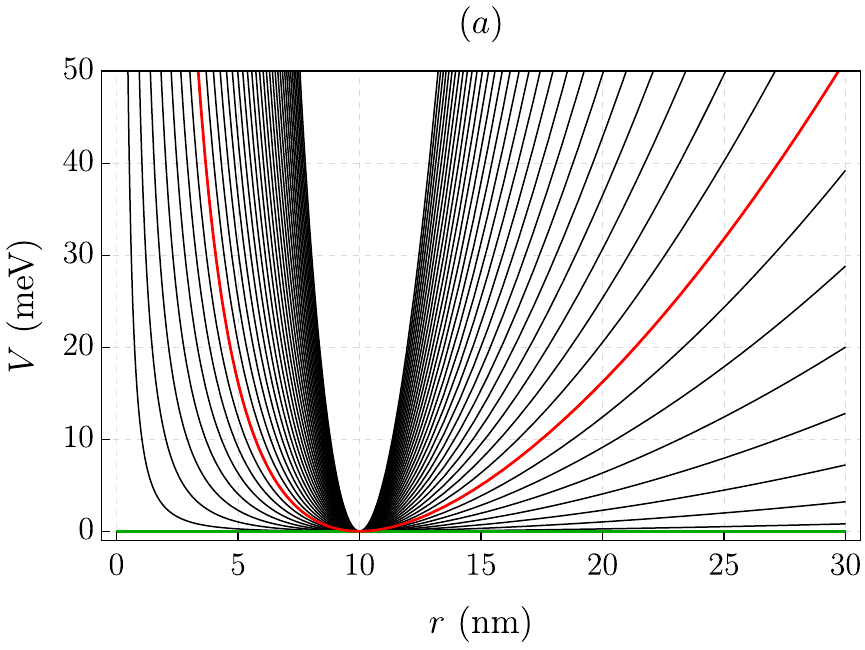}
	\includegraphics[height=5.5cm]{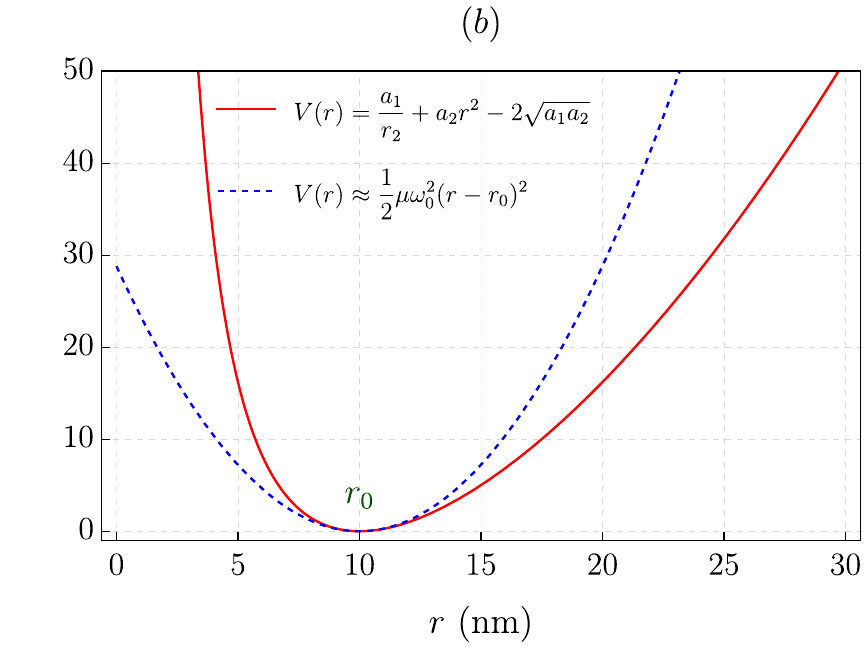}
	\caption{Profile of the Tan-Inkson potential (Equation (\ref{eq:v(r)inkson})) and the harmonic approximation potential (Equation (\ref{eq:parabola})). The parameter used for (b) is $\hbar\omega_{0}=25.6~\mathrm{meV}$ and $r_{0}=10~\mathrm{nm}$. In (a), we have $r_{0}=10~\mathrm{nm}$ with the confinement energy $\hbar\omega_{0}$ varying from $0$ to $120~\mathrm{meV}$, with a range of $3.2~\mathrm{meV}$. The green curve $\hbar\omega_{0}=0~\mathrm{meV}$ does not generate a confinement potential; the red curve in (b) is the one plotted in (a). }
	\label{fig:potencialTan-Inkson}
\end{figure}

In more general scenarios, where $a_{1}$ and $a_{2}$ are non-zero, we have a model of harmonic confining potential for a quantum ring. The resulting potential is responsible for electron confinement, restricting its radial motion. In this case, it is important to understand the behavior in the allowed region of the ring. To do this, let's study more properties of the potential (\ref{eq:v(r)inkson}). First, let's determine the minimum of this potential. By setting
\begin{align}
	\dfrac{dV}{dr} & =-\dfrac{2a_{1}}{r^{3}}+2a_{2}r=0,
\end{align}
and solving for $r$, we get
\begin{align}
	r & =\left(\dfrac{a_{1}}{a_{2}}\right)^{\frac{1}{4}}=r_{0},
\end{align}
which demonstrates that the minimum point coincides with the average radius of the ring. Expanding (\ref{eq:v(r)inkson}) in Taylor series around the point $r=r_{0}$, we have
\begin{equation}
	V(r)=V(r_{0})+\left(\dfrac{dV}{dr}\right){r=r_{0}}(r-r_{0})+\dfrac{1}{2}\left(\dfrac{d^{2}V}{dr^{2}}\right){r=r_{0}}(r-r_{0})^{2}+\dots.\label{exp}
\end{equation}
Substituting Equation (\ref{eq:v(r)inkson}) into Equation (\ref{exp}) taken at $r=r_{0}$ and their respective derivatives,
we obtain (up to third order of expansion)
\begin{equation}\label{eq:parabola}
	V(r) \approx\dfrac{1}{2}\mu\omega_{0}^{2}(r-r_{0})^{2}.
\end{equation}
Equation (\ref{eq:parabola}) tells us that, near the equilibrium point $r_{0}$, the Tan-Inkson potential behaves like a harmonic oscillator, where
$\mu$ is the electron's effective mass and $\omega_{0}=\sqrt{8a_{2}/\mu}$ characterizes the transverse confinement frequency (see Figure \ref{fig:potencialTan-Inkson}(b)).

We can find the inner radius $r_{-}$ and outer radius $r_{+}$ of the quantum ring in terms of the Fermi energy, represented by the relation \cite{SST.1996.11.1635}
\begin{equation}
	r_{\pm}=\left(\frac{V_{0}+E_{\mathrm{F}}\pm\sqrt{2E_{\mathrm{F}}V_{0}+E_{\mathrm{F}}^{2}}}{2a_{1}}\right)^{1/2},
\end{equation}
where the ring's width is determined by $\delta r=\left(r_{+}-r_{-}\right)$. 
On the other hand, the profile of the potential for an antidot is similar to that of a quantum dot viewed invertedly. While the quantum dot potential is a convex function, the antidot's is a concave function. An antidot is a region characterized by a potential barrier that repels electrons \cite{fang2023atomically}.
For a more comprehensive analysis of mesoscopic structures, we recommend consulting Refs. \cite{haug2009quantum,kalt2019semiconductor}.
\begin{figure}[tbh]
    \centering
    \includegraphics{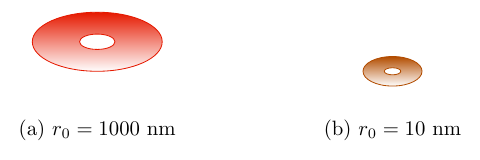}
    \caption{Representation of two quantum rings with different average radius configurations: (a) $r_0 = 10\,\mathrm{nm}$ and $\hbar\omega = 25.6\,\mathrm{meV}$, suitable for studying optical properties; (b) $r_0 = 1000\,\mathrm{nm}$ and $\hbar\omega = 0.50\,\mathrm{meV}$, used to only study the electronic properties.}
    \label{fig:ringsize}
\end{figure}

In this article, we explore two distinct configurations of QRs. The first, with a smaller average radius,  $r_0 = 10\,\mathrm{nm}$  and $\hbar\omega = 25.6\,\mathrm{meV}$, is suitable for investigating optical properties. The second, with $r_0 = 1000\,\mathrm{nm}$ and $\hbar\omega = 0.50\,\mathrm{meV}$, will be used to only obtain the graphs of the electronic properties.  These two configurations are illustrated in Figure \ref{fig:ringsize}.

In Figure \ref{fig:potencialTan-Inkson}(a), the profiles of the potential \eqref{eq:v(r)inkson} are plotted with different confinement energy values, ranging from $\hbar\omega_{0}=0~\mathrm{meV}$ to $\hbar\omega_{0}=120~\mathrm{meV}$, with a numerical interval of $\hbar\omega_{0}=6.2~\mathrm{meV}$. The graph shows that, for a given Fermi energy value, we have different confinement profiles for different $\hbar\omega_{0}$ values. For higher values, the confinement strength is greater, and we can observe that the spacing between the potential energy curves decreases. Highlighting two values seen in the graph, the green curve clearly shows that there is no confinement for the value $\hbar\omega_{0}=0~\mathrm{meV}$. The red curve is for $\hbar\omega_{0}=25.6~\mathrm{meV}$, which is analyzed in Figure \ref{fig:potencialTan-Inkson}(b) together with \eqref{eq:parabola} (dashed blue line), where it can be clearly seen that around the average ring radius $r_{0}=10~\mathrm{nm}$, the potential behaves harmonically, confirming the parabolic behavior of the Tan-Inkson potential.

The parameters used in Figure \ref{fig:potencialTan-Inkson} refer to a GaAs sample for the radial confinement potential defined by Equation (\ref{eq:v(r)inkson}), with $a_{1}=9,1022\times10^{6}\,\mathrm{meV}~\mathrm{nm}^{2}$ and $a_{2}=2,222\times10^{-5}~\mathrm{meV}\,\mathrm{nm}^{-2}$, which provide a ring with $r_{0}=10\,\mathrm{nm}$. The effective electron mass is $\mu=0.067~m_{e}$ ($m_{e}$ is the electron mass in vacuum). Near the bottom of the confinement potential, it can be well approximated by a parabolic potential with $\hbar\omega_{0}=25.6~\mathrm{meV}$. The radial asymmetry of the confining potential is because the internal repulsion is substantially stronger than the external one \cite{PE.2024.158.115898}.

\begin{figure}[tbh]
	\centering
	\includegraphics[scale=0.8]{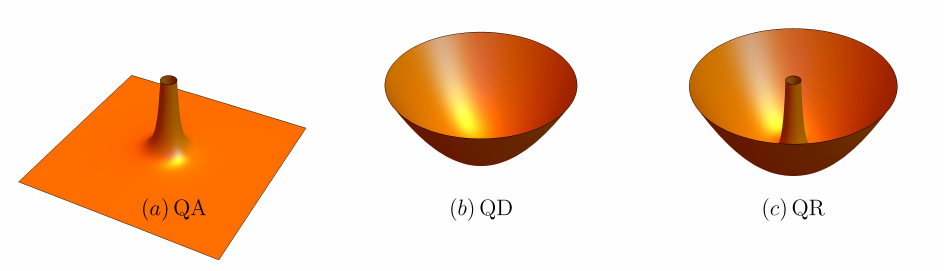}
	\caption{Three-dimensional graph of the Tan-Inkson potential (Equation (\ref{eq:v(r)inkson})). The parameter values used were $\hbar \omega_0=25.6~\mathrm{meV}$, $r_0=10~\mathrm{nm}$.}
	\label{fig:a3d}
\end{figure}

In Figure \ref{fig:a3d}, three graphs of mesoscopic potentials are presented, resulting from the modulation of the confinement coefficients of the Tan-Inkson potential (Equation \eqref{eq:v(r)inkson}), normalized for $r=\sqrt{x^{2}+y^{2}}$.
In Figure \ref{fig:a3d}(a), with $a_{1}\neq0$ and $a_{2}=0$, we have the anti-dot potential; in \ref{fig:a3d}(b), where $a_{1}=0$ and $a_{2}\neq0$, we have the quantum dot; and in \ref{fig:a3d}(c), we have the sum of the anti-dot and quantum dot potentials, thus creating a quantum ring.
In the ring graph, it is possible to observe the contribution of both the quantum dot potential and the anti-dot potential, the latter represented by the parabolic external potential wall, which confines a particle within the three-dimensional structure and prevents it from accessing regions where $r\neq0$. While the external wall acts as a confining barrier, restricting certain specific regions, the anti-dot potential is crucial for preventing the occupation of electrons in the central region, defined by the central potential barrier.
This visualization provides a broader approach to the physical problem, specifically the confinement structure.

To find the eigenfunctions and energy eigenvalues of the system, we need to solve the Schrödinger equation
\begin{align}
	&\frac{\hbar^{2}}{2\mu}\left[-\dfrac{1}{r}\dfrac{\partial}{\partial r}\left(r\dfrac{\partial}{\partial r}\right)-\dfrac{1}{r^{2}}\left(\dfrac{\partial}{\partial\varphi}+il\right)^{2}-i\dfrac{eB}{\hbar}\left(\dfrac{\partial}{\partial\varphi}+il\right)+\dfrac{e^{2}B^{2}r^{2}}{4\hbar^{2}}\right]\psi
	\notag \\&+\left(\frac{a_{1}}{r^{2}}+a_{2}\,r^{2}-V_{0}\right) \psi=E\psi.\label{shrn}
\end{align}
We use solutions of the form
\begin{equation}
	\psi(r,\varphi)=f(r)e^{-im\varphi},\label{eq:ansatz}
\end{equation}
where $ m \in \mathbb{Z} $ is the angular momentum quantum number. Substituting the ansatz (\ref{eq:ansatz}) into the Schrödinger equation (\ref{shrn}), we obtain the radial equation:
\begin{align}\label{eq:fgeome}
	f^{\prime\prime}(r)+\frac{1}{r}f^{\prime}(r)+\left(-\frac{M^{2}}{r^{2}}-\frac{\mu^{2}\omega^{2}r^{2}}{4\hbar^{2}}+\gamma^{\prime}\right)f(r)=0,
\end{align}
where
\begin{equation}
	M=\sqrt{\left(m-l\right)^{2}+\dfrac{2a_{1}\mu}{\hbar^{2}}}\label{eq:M}
\end{equation}
is the effective angular momentum,
\begin{equation}
	\omega=\sqrt{\omega_{c}^{2}+\omega_{0}^{2}}\label{eq:w}
\end{equation}
is the effective frequency,
\begin{equation}
	\omega_{c}=\dfrac{eB}{\mu}\label{eq:wc}
\end{equation}
is the cyclotron frequency,
\begin{equation}
	\lambda=\sqrt{\dfrac{\hbar}{\mu\omega}}\label{eq:lambda}	
\end{equation}
presents the effective magnetic length renormalized by the ring confinement and
\begin{align}
	\omega_0&=\sqrt{\frac{8 a_2}{\mu}},\label{eq:w0}\notag \\
	\gamma^{\prime}&=\frac{\mu\omega_{c}}{\hbar}(m-l)+\frac{2\mu}{\hbar^{2}}\left(V_{0}+E\right).
\end{align}
Equation (\ref{shrn}) is of the Kummer type whose solution is given in terms of the confluent hypergeometric function. Thus, we can express the normalized wave eigenfunctions and energy eigenvalues as
\begin{equation}\label{eq:psi}
	\psi_{nm}(r,\varphi)  =\dfrac{1}{\lambda}\left(\dfrac{\Gamma\left[n+M+1\right]}{2^{M+1}n!\left(\Gamma\left[M+1\right]\right)^{2}\pi}\right)^{\frac{1}{2}}e^{-im\varphi}
	 e^{-\frac{1}{4}\left(\frac{r}{\lambda}\right)^{2}}\left(\dfrac{r}{\lambda}\right)^{M}{}_{1}F_{1}\left(-n,M+1;\dfrac{1}{2}\left(\dfrac{r}{\lambda}\right)^{2}\right),
\end{equation}
\begin{align}
	E_{nm} & =\left(n+\dfrac{M}{2}+\dfrac{1}{2}\right)\hbar\omega-\dfrac{\left(m-l\right)}{2}\hbar\omega_{c}-\dfrac{\mu}{4}\omega_{0}^{2}r_{0}^{2},\label{eq:Enm}
\end{align}
where $n=0,1,2,3,\dots$  is the radial quantum number.
In the absence of both the AB flux ($l=0$) and the confining potential ($V(r)=0$),  the particle is subject only to the uniform magnetic field. In this case, the energy takes the form (\ref{eq:Enm})
\begin{equation}
	E_{\bar{n}}=\left(\bar{n}+\frac{1}{2}\right) \hbar \omega_c, \quad \bar{n}=n+\frac{|m|-m}{2} \geq 0,
\end{equation}
which correspond to the Landau levels \cite{PE.2024.158.115898}.
The quantum numbers $n$ and $m$ characterize the radial motion and the angular momentum, respectively. In the particular case of a circular wire, $n$ represents the index of each sub-band, and $m$ is the quantum number describing the longitudinal motion in the wire.

\begin{figure}[tbh]
	\centering
	\includegraphics[scale=0.15]{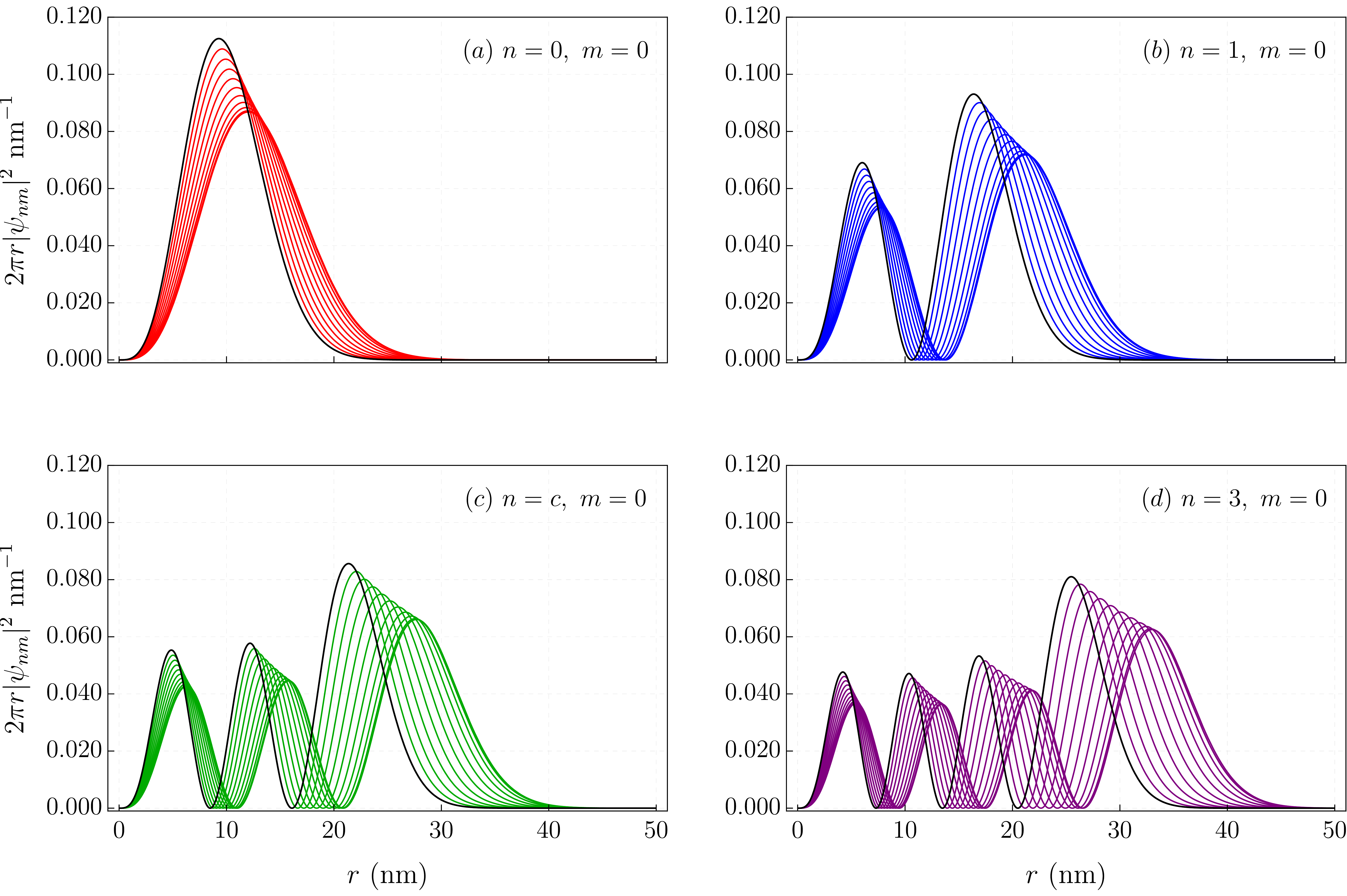}
	\caption{Probability distribution plot for the first three quantum numbers $n$. The parameter values used are $\hbar \omega_0=25.6\,\mathrm{meV}$,  $r_0=10\,\mathrm{nm}$, $\phi=0.1\,(h/e)$ and $B=0,2,4,6,\dots,20~\mathrm{T}$. The black curve in all the graphs represents the value of $B=20$.}
	\label{fig:wavefuntion}
\end{figure}

Figure \ref{fig:wavefuntion} shows the probability distribution function for the first three quantum numbers $n$
with $m=0$, where we can observe that
for the ground state $2\pi r|\psi_{00}|^{2}$ (Figure \ref{fig:wavefuntion}(a)), we have a single peak (indicating a maximum value) representing a higher probability density of finding the particle.
When we look at the first excited states (Figures \ref{fig:wavefuntion}(b), \ref{fig:wavefuntion}(c), and \ref{fig:wavefuntion}(d)), more peaks appear, but with lower probabilities. These peaks represent regions of localized probability of finding the particle. The graphs were constructed for different values of $B$, ranging from $0$ to $20~\mathrm{T}$ with an interval of $2~\mathrm{T}$ between each curve. It can be observed in each graph that the higher the value of the magnetic field (represented by the black curve in each graph), the higher the value of the probability distribution and the closer it will be to the center. This shows that the presence of the magnetic field is significant in the behavior of the particle in the quantum ring.

\begin{figure}[tbh]
	\centering
	\includegraphics[scale=0.15]{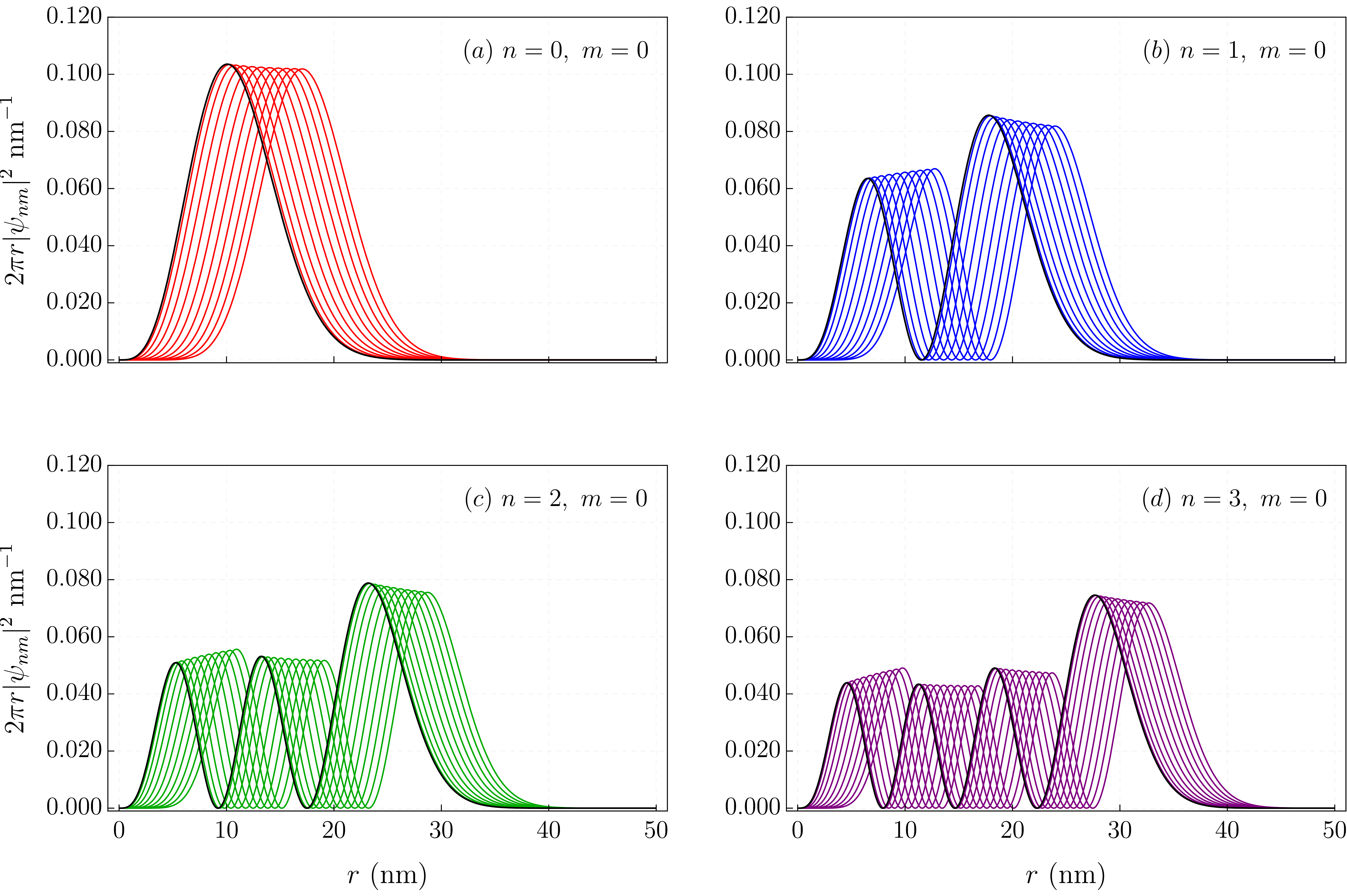}
	\caption{Probability distribution plot for the first three quantum numbers $n$. The parameter values used are $\hbar \omega_0=2.0\,\mathrm{meV}$,  $r_0=100\,\mathrm{nm}$, $\phi=0,0.4,0.8,1.2,\dots,4~(h/e)$ and $B=0~\mathrm{T}$. The black curve in all the graphs represents the value of $\phi=0$.}
	\label{fig:wavefuntionphi}
\end{figure}

The same analysis was performed for Figure \ref{fig:wavefuntionphi}, which shows the probability distribution from the ground state to the next excited states,
but now varying the value of the AB flux, with the value of $\phi$ ranging from $0$ to $4~(h/e)$ with an interval of $0.4~(h/e)$. In this set of graphs, we can observe that the contribution of $\phi$ moves the probability density away from the center of the ring, represented by the black curves further away from $0$ on the $r$ axis. This also shows the influence of the AB flux on the quantum system.

\begin{figure}[tbh]
	\centering
	\includegraphics[scale=0.55]{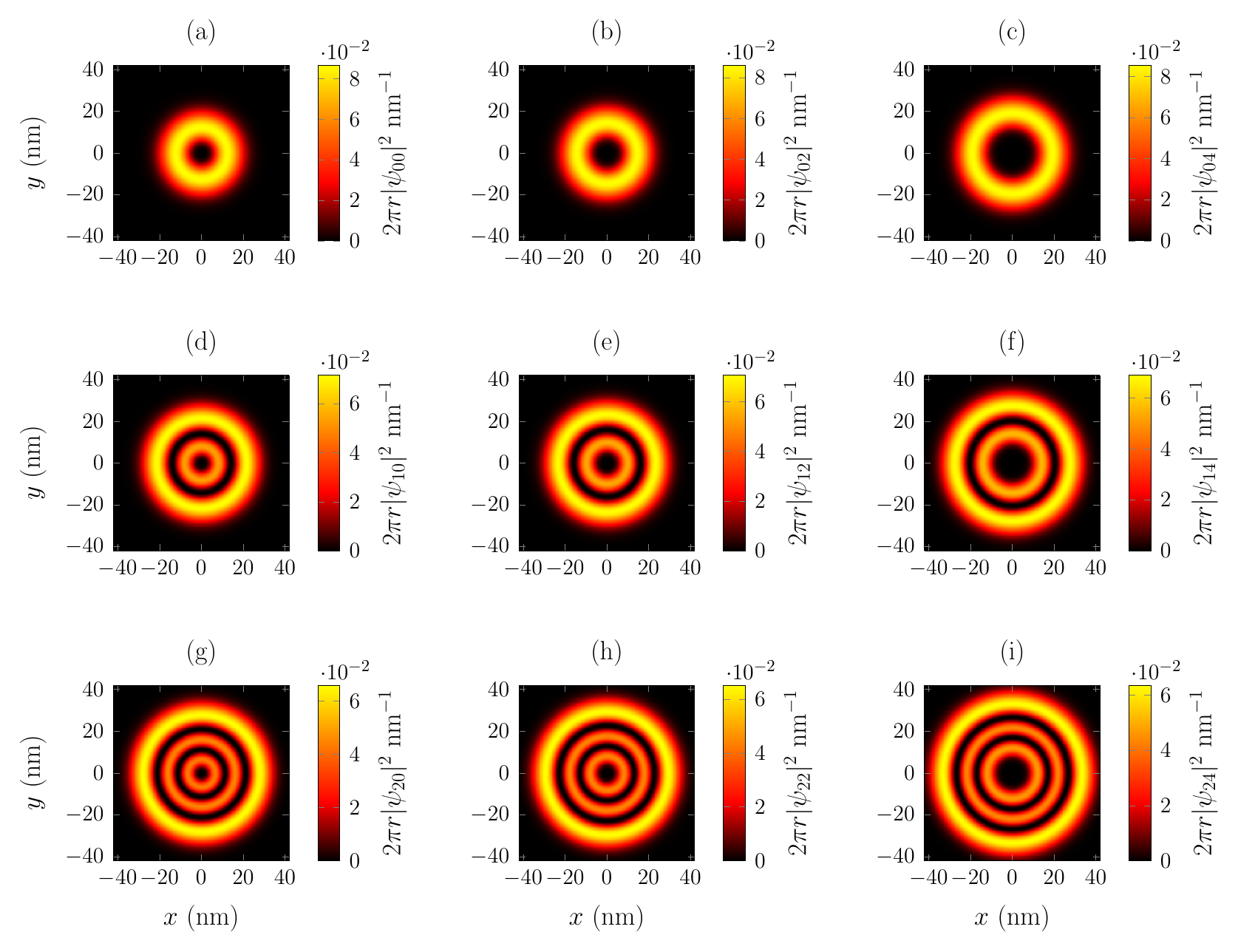}\\    
	\caption{ Normalized probability density in the $xy$ plane for some values of $n$ and $m$. The parameter values used are $r_{0}=10~\mathrm{nm}$, $\hbar\omega_{0}=25.6~\mathrm{meV}$.}
	\label{fig:densityplot}
\end{figure}

A very intuitive way to visualize the annular geometry of the QRs is by constructing the probability distribution function in the $xy$ plane of the density function. Figure \ref{fig:densityplot} shows a set of density plots corresponding to the solution given by Equation (\ref{eq:psi}) in the transverse plane normalized by $r=\sqrt{x^{2}+y^{2}}$ in nanometers, for different quantum numbers $n$ and $m$. 

For $m=0$, the structure shows a core at the origin $(x=0,\,y=0)$. In the ground state, the probability of finding the electron is higher near the center of the ring, as illustrated in the first plot of Figure \ref{fig:densityplot}. For a given value of $n$, the probability of finding the particle shifts away from the center as $m$ increases, which is represented by the widening of the maximum away from the center, showing the presence of a centrifugal potential associated with $m$. Each plot describes a probability density associated with each state of $2\pi r|\psi_{nm}|^{2}$, making the analysis more intuitive about the behavior of the particle in the quantum ring as the quantum numbers $n$ and $m$ vary. It is important to note that the plots were generated for even values of $m$, without any specific reason for this selection. In fact, these values could be arbitrary as long as they obey the condition $m = 0, \pm 1, \pm 2, \cdots$. The choice to limit to even values of $m$ was made to highlight the observed effects when varying $m$ more clearly.

The analysis of electronic properties is of great importance to help us understand how states behave when subjected to the effects of a magnetic field. For a better understanding of the model that will be presented in the following, in this section, we reproduce some results from Ref. \cite{SST.1996.11.1635}. Initially, we assume $\omega_{c}=0$ (absence of uniform magnetic field) in Equation (\ref{eq:Enm}), such that the energy spectrum of the $2\mathrm{D}$ ring is written as
\begin{equation}
	E_{nm} =\left(n+\dfrac{1}{2}\sqrt{\left(m-l\right)^{2}+\dfrac{2a_{1}\mu}{\hbar^{2}}}+\dfrac{1}{2}\right)\hbar\omega-\dfrac{\mu}{4}\omega_{0}^{2}r_{0}^{2}.\label{expc}
\end{equation}
Equation (\ref{expc}) reveals that states with $m=0$ are all symmetric for all energy subbands. The effect of $\omega_{c} \neq 0$ produces a non-parabolic dispersion that breaks the present symmetry, revealing the existence of a centrifugal potential responsible for such dispersion. This potential is the reason why different states of the same subband have different radial wave functions \cite{SST.1996.11.1635}.

\begin{figure}[tbh]
\centering
\includegraphics[scale=0.65]{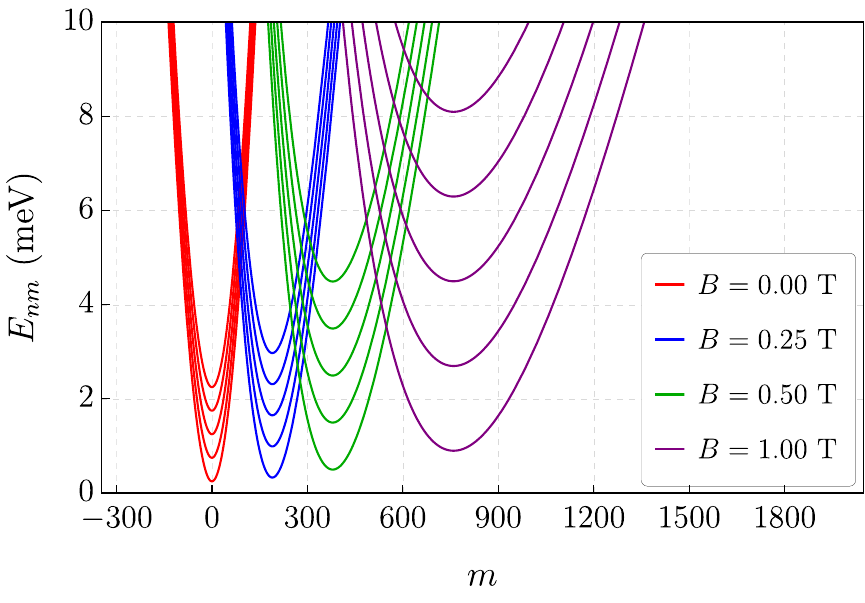}
\caption{ Eigenenergies of the $2\mathrm{D}$ ring as a function of the quantum number $m$ for the first five lowest subbands $(n = 0,\,1,\,2,\,3,\,4)$ at different magnetic field intensities. The parameter values used are $\hbar\omega_{0}=0.50$\,meV, $\phi=0\,(h/e)$, and $r_{0}=1000$~nm. This figure is a modified adaptation from the original work of Tan-Inkson, developed for a different quantum ring configuration \cite{SST.1996.11.1635}.}
\label{fig:EnergiaSubbanda}
\end{figure}

Figure \ref{fig:EnergiaSubbanda} shows the graph of energy levels as a function of the quantum number $m$ of a $2\mathrm{D}$ ring made from a GaAs heterostructure for the five lowest subbands, assuming different values of the magnetic field $B$.

For the case where $B=0$, the subbands exhibit a parabolic shape centered around $m=0$. As the magnetic field value increases, a significant dispersion can be observed, evidenced by the increase in energy at the subband minima and the increase in spacing between them \cite{PE.2024.158.115898}. The energy subbands are directly related to the standing waves that extend throughout the $2\mathrm{D}$ ring. Furthermore, the energy is periodically modulated with the magnetic field, known in the literature as AB oscillations \cite{SST.1996.11.1635}.

\begin{figure}[tbh]
	\centering
	\includegraphics[scale=0.65]{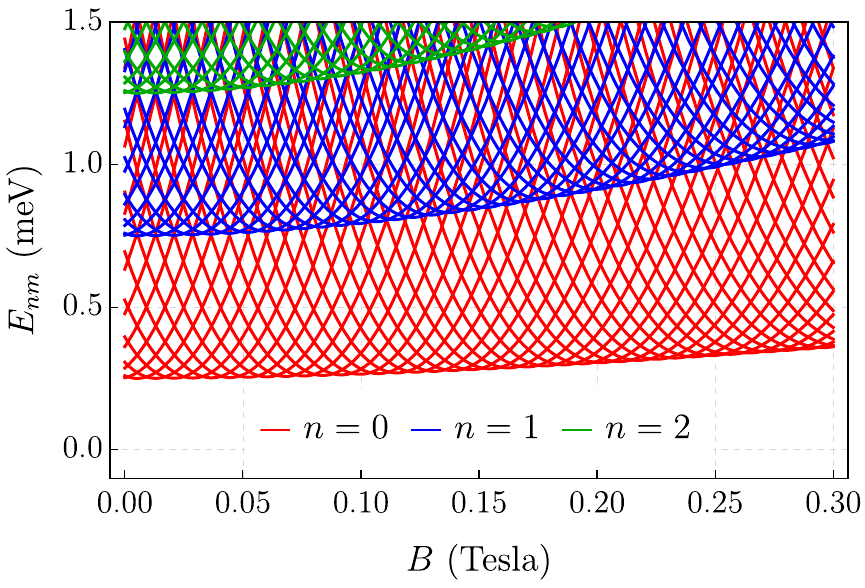}\
	\caption{ The three lowest subbands, where the parameter values used are $\hbar\omega_{0}=0.50$\,meV, $\phi=0\, (h/e)$, and $r_{0}=1000$\,nm. This figure is a modified adaptation from the original work of Tan-Inkson, developed for a different quantum ring configuration \cite{SST.1996.11.1635}.}
	\label{fig:EnergiaSubbandacampomagnetico}
\end{figure}

The degeneracy of states for the quantum ring can be observed in Figure \ref{fig:EnergiaSubbandacampomagnetico}, where we outline the energy spectrum as a function of the magnetic field. Each line represents a distinct quantum number $m$, while lines of the same color represent the same quantum number $n$, which characterizes a subband of the system. In other words, each curve represents a distinct state $\psi_{nm}$ of the system. It can be observed that the energy spectrum as a function of the applied magnetic field is aperiodic, especially when more than one subband is occupied \cite{SST.1996.11.1635}.

\begin{figure}[tbh]
\centering
\includegraphics[scale=0.6]{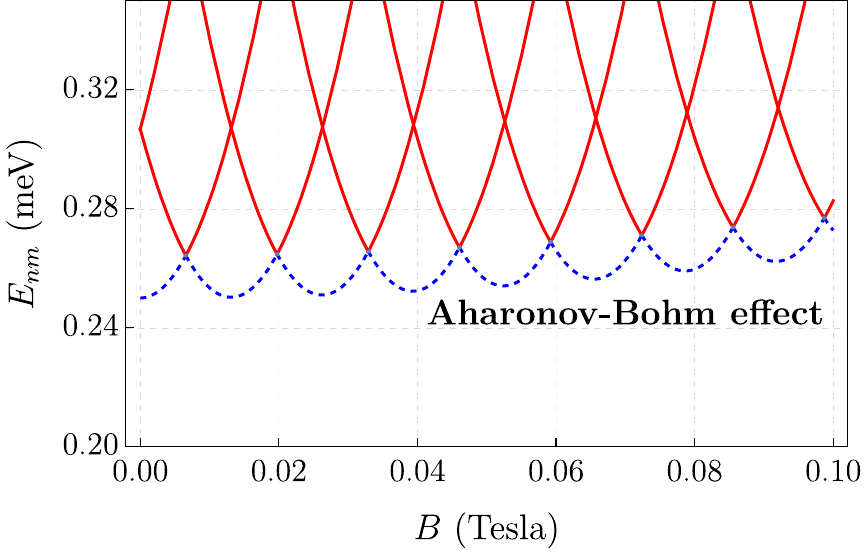}\
\caption{ Aharonov-Bohm oscillation for the lowest sub-band $n=0$. The dashed curves indicate the ground state changing its quantum number $m$ by one unit. The parameter values used were the same as in Figure \ref{fig:EnergiaSubbanda}.}
\label{fig:efeAB}
\end{figure}

The graph in Figure \ref{fig:efeAB} is an enlargement of Figure \ref{fig:EnergiaSubbandacampomagnetico}, where the energy spectrum in the ground state (parts of the curves in red color) is schematically shown for an electron exhibiting oscillations as a function of the magnetic field (AB Oscillations).
It can be observed that at each energy maximum, there is a transition in the ground state. At this intersection point, the quantum number of angular momentum $m$ of the ground state changes by one unit, i.e., there is an exchange of angular momentum between the states \cite{vladimirphysics}. These features in electronic properties, especially the energy spectrum of the quantum ring in terms of magnetic field and flux, are of scientific interest, particularly in investigating the AB effect. 

\begin{figure}tbh]
\centering
\includegraphics[scale=0.65]{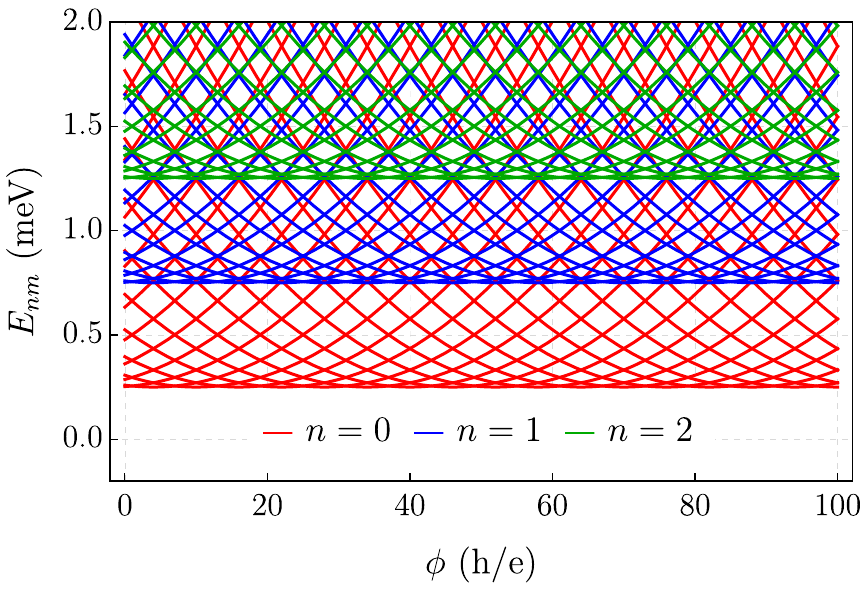}
\caption{ The three lowest subbands, where the parameter values used are $\hbar\omega_{0}=0.50~\mathrm{meV}$, $B=0$ T, and $r_{0}=1000~\mathrm{nm}$. This figure is a modified adaptation from the original work of Tan-Inkson, developed for a different quantum ring configuration \cite{SST.1996.11.1635}.}
\label{fig:EnergiaSubbandafluxomagnetico}
\end{figure}

Figure \ref{fig:EnergiaSubbandafluxomagnetico} presents the energy eigenvalues as a function of the magnetic flux for the three lowest subbands. It is observed that the energy spectrum exhibits periodicity concerning the AB flux, with a period of $\phi_{0}$ \cite{SST.1996.11.1635}. Furthermore, the graph shows that energy levels can occupy multiple subbands. Additionally, as shown in Figure \ref{fig:EnergiaSubbandacampomagnetico}, the energy minima of each subband increase for larger values of $n$.

In the following sections, we will investigate this phenomenon during optical transitions observed in semiconductor materials.

\section{Photoionization Process for the 2D Quantum Ring Model}\label{sec3}

In this section, we investigate the Photoionization Cross Section (PCS) of a two-dimensional quantum ring composed of GaAs. We also show how the AB flux manifests when some model parameters are varied. This model was previously studied by Xie \cite{SM.2013.58.94}, who investigated the effects of AB flux on the PCS in two-dimensional QRs. However, Xie's approach does not entirely follow the Tan-Inkson model. Here, we will rigorously follow the Tan-Inkson derivation, as presented in Section \ref{sec2}, to derive the solutions with specific quantum transitions and parameters for application to the PCS formula. Therefore, we can conclude that the contribution in this section is largely novel to the literature.

The PCS represents the probability of a bound electron being liberated by appropriate radiation with energy $\hbar \omega$ of a particular frequency. Its magnitude is profoundly influenced by the confinement potential and the photon energy \cite{articleTshipa,tshipa2021photoionization}. 
The dependence of excitation energy in the photoionization process is derived from Fermi's golden rule using the well-known dipole approximation written as \cite{lax1954,lax1952franck,PRB.2008.77.045317}
\begin{align}
	\sigma \left( \hbar \omega \right) =C_{n_{r}}\hbar \omega
	\sum\limits_{f} &\left\vert \left\langle \psi_{n,m}\left\vert e\boldsymbol{r}
	\right\vert \psi_{n^{\prime },m^{\prime}}\right\rangle \right\vert
	^{2} \delta \left({E}_{n,m}^{\left(f\right)}-{E}_{n,m}^{\left(i\right)
	}-\hbar \omega \right),\label{cs}
\end{align}
with
\begin{equation}
	\delta \left( E_{n,m}^{\left( f\right) }-E_{n,m}^{\left( i\right) }-\hbar
	\omega \right) =\frac{1}{\pi }\frac{\hbar \Gamma _{f}}{\left(
		E_{n,m}^{\left( f\right) }-E_{n,m}^{\left( i\right) }-\hbar \omega \right)
		^{2}+\left( \hbar \Gamma _{f}\right) ^{2}},\label{eq:lzr}
\end{equation}
where $\Gamma_{f i}$ represents the relaxation rate between the initial and final states. The constant $C_{n_{r}}$ in Equation (\ref{cs}) takes into account the electrical and optical properties of the semiconductor material, being given by
\begin{equation}\label{eq:cnr}
	C_{n_{r}}=\left(\frac{\xi_{\text{eff}}}{\xi_{0}}\right)^{2}\frac{n_{r}}{\epsilon}\frac{4\pi^{2}}{3}\alpha_{\text{fs}},
\end{equation}
where $n_{r}$ is the refractive index, $\epsilon$ is the optical dielectric constant, $\alpha_{\mathrm{fs}}=e^{2}/\hbar c$ represents the fine structure constant, ${\xi_{eff}}/{\xi_{0}}$ is the ratio of the effective electric field, with $\xi_{eff}$ representing the amplitude of the electric field of the incident radiation and $\xi_{0}$ the average field in the semiconductor. Furthermore, this rate is related to the time $\tau_{f i}$ of the transition $|\psi_{nm}^{(i)}\rangle \rightarrow|\psi_{nm}^{(f)}\rangle$, with $\Gamma_{f i}=1 / \tau_{f i}$ \cite{EPJB.2021.94.129}. This ratio does not affect the shape of the PCS, and, at present, no experimental data is available in the literature for comparison \cite{SM.2013.58.94}. Thus, this ratio is often considered equal to $1$, and we will not explicitly include it in our results. See Ref. \cite{ridley2013quantum} for more information on this issue. In Equation (\ref{cs}), we can define $\left\langle\psi_{n, m}^{(i)}|\boldsymbol{r}| \psi_{n, m}^{(f)}\right\rangle \equiv \mathcal{M}_{i f}$ as the matrix element between the initial and final states of the dipole moment. The functions $\psi_{n, m}^{(i)}$ and $\psi_{n, m}^{(f)}$ symbolize the wave functions of the initial and final states, while ${E}_{n, m}^{(f)}$ and ${E}_{n, m}^{(i)}$ correspond to the respective energy eigenvalues of the transition. 
The quantity $\mathcal{M}_{i f}$ can be calculated using the solution (\ref{eq:psi}). Hence, we have
\begin{align}
	\mathcal{M}_{if}& =\frac{1}{2\pi \lambda _{0}^{2}}\left[ \frac{\Gamma \left(
		n+M+1\right) }{n!\left( \Gamma \left( M+1\right) \right) ^{2}}\right] ^{
		\frac{1}{2}}\left[ \frac{\Gamma \left( n^{\prime }+M^{\prime }+1\right) }{
		n^{\prime }!\left( \Gamma \left( M^{\prime }+1\right) \right) ^{2}}\right] ^{
		\frac{1}{2}}  \int_{0}^{\infty }\int_{0}^{2\pi }e^{i\left( m^{\prime }-m\right)
		\varphi }\notag \\
	& \times \, _{1}F_{1}\left( -n,M+1,\frac{r^{2}}{2\lambda _{0}^{2}}\right) \cos \varphi \,
	e^{-\frac{r^{2}}{2\lambda _{0}^{2}}}\left( \frac{r^{2}}{2\lambda
		_{0}^{2}}\right) ^{\zeta }\, _{1}F_{1}\left( -n^{\prime },M^{\prime }+1,\frac{r^{2}}{
		2\lambda _{0}^{2}}\right) r^{2}drd\varphi, 
\end{align}
with $\zeta =\left( M+M^{\prime }\right) /2$. Defining the new variable $
z=r^{2}/2\lambda _{0}^{2}$, we write
\begin{equation}
	\mathcal{M}_{if}=c_{nm}^{\prime }\int_{0}^{2\pi }e^{i\left( m^{\prime
		}-m\right) \varphi }\cos \varphi d\varphi \int_{0}^{\infty }e^{-z}z^{\zeta +
		\frac{1}{2}}  
	\,_{1}F_{1}\left( -n^{\prime },M^{\prime }+1,z\right)
	\,_{1}F_{1}\left( -n,M+1,z\right) dz,
\end{equation}
where
\begin{equation}
	c_{nm}^{\prime }=\frac{\lambda _{0}}{\sqrt{2}}\left[ \frac{\Gamma \left(
		n^{\prime }+M^{\prime }+1\right) }{n^{\prime }!\left( \Gamma \left(
		M^{\prime }+1\right) \right) ^{2}}\right] ^{\frac{1}{2}}\left[ \frac{\Gamma
		\left( n+M+1\right) }{n!\left( \Gamma \left( M+1\right) \right) ^{2}}\right]
	^{\frac{1}{2}}.
\end{equation}
Using the identity $2\cos \varphi =e^{i\varphi }+e^{-i\varphi }$, the integration over the coordinate $\varphi $ results in
\begin{equation}
	\int_{0}^{2\pi }e^{i\left( m^{\prime }-m\right) \varphi }\cos \varphi\,
	d\varphi =\pi \left( \delta _{m^{\prime },m-1}+\delta _{m^{\prime
		},m+1}\right) .
\end{equation}
Thus, we obtain
\begin{equation}
\mathcal{M}_{if} =c_{nm}^{\prime
}\left( \delta_{m^{\prime },m-1}+\delta _{m^{\prime },m+1}\right)
\int_{0}^{2\pi }e^{-z} dz 
\,z^{\zeta +\frac{1}{2}}\,_{1}F_{1}\left(-n^{\prime},M^{\prime
}+1,z\right)\,_{1}F_{1}\left( -n,M+1,z\right) dz.
\end{equation}
From this result, we see that the non-zero elements of the matrix follow the selection rule $\Delta m= \pm 1$ \cite{Pereira_2024}.

To obtain more consistent results, we conduct a numerical investigation exploring relevant data readily available in the literature, particularly for the case of GaAs two-dimensional QRs. From now on, we present a detailed numerical analysis of our results. All numerical calculations were performed using the following parameters: $\epsilon=13.1$, $\mu = 0.067 \,\mu_e$, where $\mu_e=0.511\,\text{MeV}/c^{2}$, $\epsilon_0=8.854\times10^{-12}\,\text{F/m}$, $\mu_0=4\pi\times10^{-7}\,\text{Tm/A}$, and $c=2.99\times10^{8}\,m/s$ \cite{PM.2019.99.2457}. We also consider the specific values: $n_{r}=3.15$, $\alpha_{\text{fs}}=1/137$, ${\xi_{\text{eff}}}/{\xi_{0}}=1$ \cite{PBB.2008.77.045317}, and also use $\hbar\Gamma_{f}=0.1 \,\text{meV}$ \cite{EPJB.2021.94.129}.

\begin{figure}[tbh]
	\centering
	\includegraphics[scale=0.9]{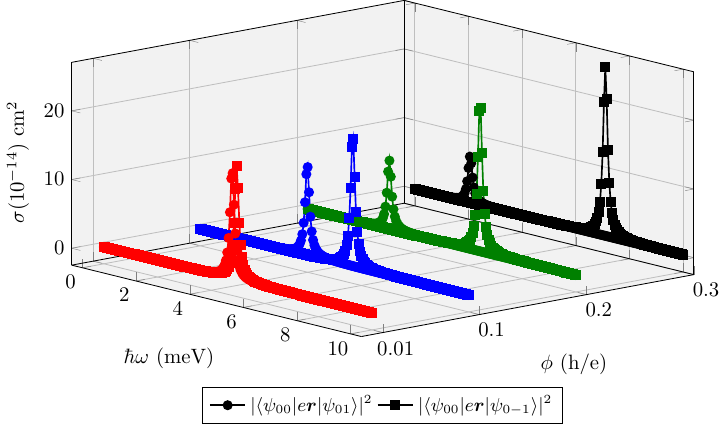}
	\caption{ Photoionization cross-section as a function of incident photon energy for different AB flux values. The parameter values used were $\hbar \omega_0=25.6~\mathrm{meV}$, $r_0=10~\mathrm{nm}$, and $B=10~\mathrm{T}$.}
	\label{fig:dm}
\end{figure}

Figure \ref{fig:dm} shows the PCS, from Equation (\ref{cs}), as a function of $\hbar \omega$ (the incident photon energy) for different values of $\phi$. 
We can see that, as the AB flux varies, the amplitudes of the PCS peaks change proportionally for two optical transitions starting from the ground state to adjacent states, following the selection rule. The reason for such behavior is that the wave function depends on the applied AB flux. Furthermore, it can be observed that the peak position undergoes a shift towards lower energies (redshift) with the increase in AB flux, where the optical transition occurs from the state $\psi_{00}$ to the state $\psi_{01}$.
The opposite effect is observed when the transition occurs from the state $\psi_{00}$ to $\psi_{0-1}$, where the PCS peaks shift to higher energy values as the AB flux increases (blue shift). The explanation for this behavior lies in the energy difference between the two states involved in the optical transition. For example, the redshift indicates that the energy difference between the two considered quantum states will decrease as the AB flux increases. On the other hand, the opposite occurs with the blue shift, which is graphically represented by the separation between the peaks of the two transitions as the AB flux increases, such that for $\phi=0$ the curves of the two transitions overlap. To demonstrate this effect more explicitly, we consider transitions between states with $\phi=0.1\,{(h/e)}$ and $\phi=0.2\,{(h/e)}$, respectively, given by
\begin{align}
	E^{(1)}_{fi}=E_{f}-E_{i}\approx 3.982~\mathrm{meV}\longrightarrow |\langle\psi_{00}|\psi_{01}\rangle|^{2},\label{t1}\\
	E^{(2)}_{fi}=E_{f}-E_{i}\approx 3.042~\mathrm{meV}\longrightarrow |\langle\psi_{00}|\psi_{01}\rangle|^{2},\label{t2}
\end{align}
such that 
\begin{equation}
	E^{(1)}_{fi}>E^{(2)}_{fi}.
\end{equation}
The transitions (\ref{t1}) and (\ref{t2}) were obtained using Equation (\ref{eq:Enm}) and subsequently used to compute the Lorentzian (\ref{eq:lzr}). It can be noted that the energy difference for the same transition decreases with the increase in magnetic flux, as explained earlier. For the other transitions, we have
\begin{align}
	E^{(1)}_{fi}=E_{f}-E_{i}\approx 5.681~\mathrm{meV}\longrightarrow |\langle\psi_{00}|\psi_{0-1}\rangle|^{2},\\
	E^{(2)}_{fi}=E_{f}-E_{i}\approx 6.426~\mathrm{meV}\longrightarrow |\langle\psi_{00}|\psi_{0-1}\rangle|^{2}.
\end{align}
In this case, the inequality is inverted, i.e.,
\begin{equation}
	E^{(1)}_{fi}<E^{(2)}_{fi}.
\end{equation}

\begin{figure}tbh]
	\centering
	\includegraphics[scale=0.9]{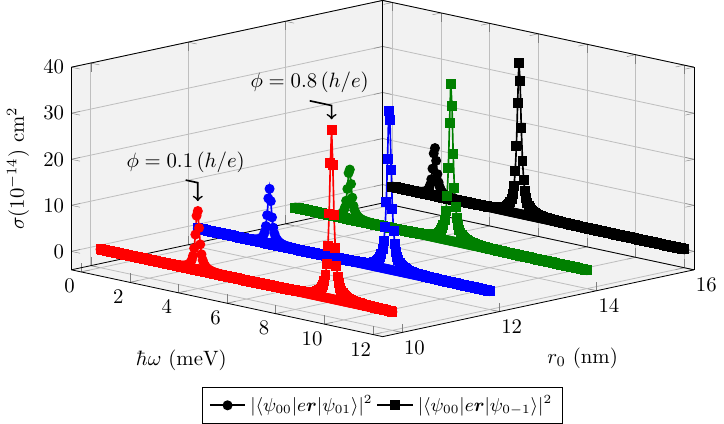}
	\caption{Photoionization cross-section as a function of incident photon energy for different average radius. The parameter values used were $\hbar \omega_0=25.6~\mathrm{meV}$ and $B=10~\mathrm{T}$.}
	\label{fig:r0}
\end{figure}

On the other hand, in Figure \ref{fig:r0}, an interesting aspect is shown. As $r_{0}$ decreases, the magnitude of the peak increases, and its position shifts to higher energy values.
The physical reason for this behavior is that there is a higher probability of the optical transition occurring between the states $\psi_{i}$ and $\psi_{f}$ in a narrower quantum ring \cite{Mingge2013}. In the figure, it is possible to observe that the PCS curves were constructed by assigning a fixed value of $\phi$ for the two optical transitions. The smaller peaks correspond to an AB flux value of $0.1\,(h/e)$, represented by curves with filled circles. By making a small variation in $\phi$, assigning it a value of $0.8\,(h/e)$, the peaks (represented by curves with filled squares) increase to higher PCS values. It is also noted that as the radius increases, the distance between the peaks of the two transitions for the same value of $\phi$ decreases. This demonstrates the effect of a change in AB flux on the probability of an optical transition occurring in the quantum ring.

\begin{figure}[tbh]
	\centering
	\includegraphics[scale=0.9]{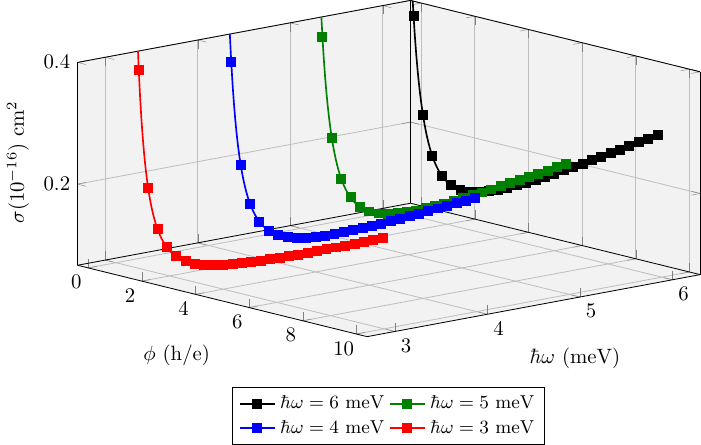}
	\caption{ Photoionization cross-section as a function of $\phi$ for different incident energy photons. The parameter values are the same as in Figure \ref{fig:dm}.}
	\label{fig:pcsVSflux}
\end{figure}

Since one of the main objectives in this section is to investigate the Aharonov–Bohm effect present in the PCS, in Figure \ref{fig:pcsVSflux}, we plot the PCS curve profile as a function of $\phi$ for different values of $\hbar \omega$. As expected, the curve exhibits a non-monotonic character, which agrees with what is presented in the literature \cite{SM.2013.58.94,Mingge2013}. It can be noted that the probability of the transition occurring is higher for more energetic photons.

\section{2D Quantum Ring under Effects of Rotation}\label{sec4}

In this section, we study the non-relativistic quantum mechanics of a particle confined to move in a 2D quantum ring under the effects of rotation. We also consider the presence of a magnetic flux tube (AB effect) and a uniform magnetic field. We do not consider the particle's spin effects in our approach. The reason for this is that the particle is confined only to the ring defined by the confining potential (\ref{eq:v(r)inkson}), and the spin of the particle interacts with the field inside the solenoid only if it accesses the $r=0$ region (an idealized null radius flow tube). Our main goal is to determine the eigenfunctions and eigenvalues of energy and then compare them with the results obtained in Section \ref{sec2}, where we consider the case of quantum rings in the absence of rotation.

\begin{figure}[tbh]
	\centering
	\includegraphics[scale=1]{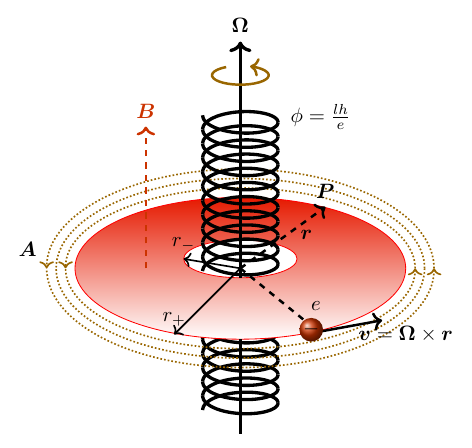}
	\caption{Illustration of an electron in a rotating frame with angular velocity $\boldsymbol{\Omega}$ around $\boldsymbol{\hat{z}}$. Adapted from Ref. \cite{Pereira_2023}.}
	\label{fig:anelcomgiro}
\end{figure}

The description of a particle model in a rotating frame is usually done via minimal coupling directly into the system's Hamiltonian. Figure \ref{fig:anelcomgiro} schematically depicts a particle confined to a quantum ring in a rotating frame. The ring is also subject to a uniform magnetic field and in the presence of the AB effect. Electrons are confined to the region bounded by the inner radius $r_{-}$ and outer radius $r_{+}$ (the fully shaded area in Figure \ref{fig:anelcomgiro}).
The dynamics of this system take into account the effects of rotation and can be included in the equation of motion through the substitution
\cite{rizzi2004relativity}
\begin{equation}
	p^{\tau}\rightarrow p^{\tau}-\mu\mathcal{A}^{\tau},\;\; (\tau =0,1,2,3),\label{mc}
\end{equation}
where $\mu$ is the particle's mass and $\mathcal{A}^{\tau}$ is the gauge field for the rotating frame, defined as 
\begin{equation}
	\mathcal{A}^{\tau}=(\mathcal{A}_{0},\boldsymbol{\mathcal{A}})=\left(-\frac{1}{2} v^2, \boldsymbol{v}\right),
\end{equation}
where $\boldsymbol{v}$ is the particle's velocity vector, which is related to the angular velocity $\boldsymbol{\Omega}$ as $\boldsymbol{v}=\boldsymbol{\Omega}\times\boldsymbol{r}$ (see Figure \ref{fig:anelcomgiro}), where $\boldsymbol{r}$ is the position vector from the center of the ring to where the particle is located. Additionally, there is an interaction of the system with a uniform magnetic field $\boldsymbol{B}$, which, along with $\boldsymbol{\Omega}$, are oriented in the positive direction of $\boldsymbol{\hat{z}}$, and strictly follows the right-hand rule. On the other hand, if it is oriented in the opposite direction, it will exhibit different physical behaviors. These characteristics will be discussed further. 

The interaction of a charged particle with the magnetic field is described through the coupling (\ref{mc}). In this case, the four-potential is defined as $A^{\tau}=(A_{0},\boldsymbol{A})$. The Hamiltonian describing the motion of the particle in the rotating frame is written as \cite{AoP.2020.419.168229,DANTAS201511,rizzi2004relativity,Universe.2024.10.389,PRA.2022.106.022211,EPJP.2021.136.920}
\begin{equation}
	H_{\Omega}=\frac{1}{2 \mu}\left(\boldsymbol{p}-e \boldsymbol{A}\right)^2-\boldsymbol{\Omega} \cdot \boldsymbol{L}+V(\boldsymbol{r}),\label{hhm}
\end{equation}
where $\boldsymbol{L}=\boldsymbol{r}\times\boldsymbol{p}$ is the particle's angular momentum and $V(\boldsymbol{r})$ is the scalar potential defined in Equation (\ref{eq:v(r)inkson}) responsible for the particle's confinement, and $\boldsymbol{A}$ is the vector potential given in Equation (\ref{pv}). For the sake of simplicity, in the present model, we consider the idealized model where the sample rotates with a constant angular velocity ($\Dot{\boldsymbol{\Omega}}=0$) around the $\boldsymbol{z}$ axis, i.e., $\boldsymbol{\Omega}=\Omega\boldsymbol{\hat{z}}$, with $\boldsymbol{r}=\left(r_{0},0,0\right)$ being a vector locating a fixed point $\boldsymbol{P}$ as shown in Figure \ref{fig:anelcomgiro}.

Considering the fields and potentials as defined earlier,
the Hamiltonian for the system is written as
\begin{equation}\label{eq:hrotti}
	\mathcal{H}_{\Omega}=\frac{1}{2\mu}\left(\boldsymbol{p}-e\boldsymbol{A}-\mu\boldsymbol{\Omega}\times\boldsymbol{r}\right)^{2}-\frac{1}{2}\mu\left(\boldsymbol{\Omega}\times\boldsymbol{r}\right)^{2}+V(r),
\end{equation}
and the Schrödinger equation to be solved is
\begin{align}\label{eq:SchrodingerRot}
	\frac{1}{2 \mu}\left(\boldsymbol{p}-e\boldsymbol{A}-\boldsymbol{A}_{\Omega}\right)^2 \Psi-\frac{1}{2} \mu \boldsymbol{A}_{\Omega}^2 \Psi+\frac{a_1}{r^2} \psi+a_2 r^2 \Psi =\left(E+V_0\right) \Psi,
\end{align}
where, for convenience, we have defined the term $\boldsymbol{A}_{\Omega}=\boldsymbol{\Omega} \times \boldsymbol{r}$, which originates from the minimal coupling and can be understood as a gauge field (effective vector potential) associated with the non-inertial effect of the rotating frame. 

In the same way as in Equation (\ref{eq:ansatz}), we use eigenfunctions in the form
\begin{equation}
	\Psi (r,\varphi)=R(r)\,e^{im\varphi},\label{eq:ansatzr}
\end{equation}
where $m=0,\pm1,\pm2,\pm3,\cdots$ is the quantum angular momentum number, and the function $R(r)$ satisfying the radial differential equation
\begin{align}
	R^{\prime\prime}+\dfrac{1}{r}R^{\prime}+\left[-\frac{L^{2}}{r^{2}}-\frac{\mu^{2}\varpi^{2}r^{2}}{4\hbar^{2}}+\gamma^{\prime}\right]R & =0,\label{eq:radialrr}
\end{align}
with
\begin{equation}
	\gamma^{\prime} =\dfrac{\mu\omega^{*}}{\hbar}\left(m-l\right)+\frac{2\mu}{\hbar^{2}}\left(V_{0}+\mathcal{E}\right),
\end{equation}
where we have defined the quantities
\begin{align}
	\omega_{c}^{2} & =\left(\dfrac{eB}{\mu}\right)^{2},\\
	\omega_{0}^{2} & =\dfrac{8a_{2}}{\mu},\\
	\varpi^{2} & =\omega_{c}^{2}+4\Omega\omega_{c}+\omega_{0}^{2},\label{eq:varpi}\\
	\omega^{*} & =\omega_{c}+2\Omega,\\
	L^{2} & =\left(m-l\right)^{2}+\frac{2\mu a_{1}}{\hbar^{2}}.
\end{align}
The energy eigenvalues and normalized wave functions of equation (\ref{eq:radialrr}) are
\begin{equation}
	\mathcal{E}_{nm}=\left(n+\dfrac{L}{2}+\dfrac{1}{2}\right)\hbar\varpi-\dfrac{1}{2}\left(m-l\right)\hbar\omega^{*}-\dfrac{\mu}{4}\omega_{0}^{2}r_{0}^{2},\label{eq:energirrnm}
\end{equation}
\begin{align}\label{eq:upsilon}
	\Psi_{n,m}(r,\varphi) & =\dfrac{1}{\lambda_{0}}\sqrt{\dfrac{\Gamma\left(L+n+1\right)}{2^{L+1}\pi n!\left[\Gamma\left(L+1\right)\right]^{2}}}\left(\frac{r}{\lambda_{0}}\right)^{L}e^{-\frac{1}{4}\left(\frac{r}{\lambda_{0}}\right)^{2}}\,_{1}F_{1}\left[-n;L+1;\frac{r^{2}}{2\lambda_{0}^{2}}\right]e^{im\varphi},
\end{align}
with
\begin{equation}
	\lambda_{0}=\sqrt{\dfrac{\hbar}{\mu\varpi}}
\end{equation}
being the new effective magnetic length renormalized by the ring confinement and the rotation. 
\begin{figure}[tbh]
\centering
\includegraphics[scale=0.15]{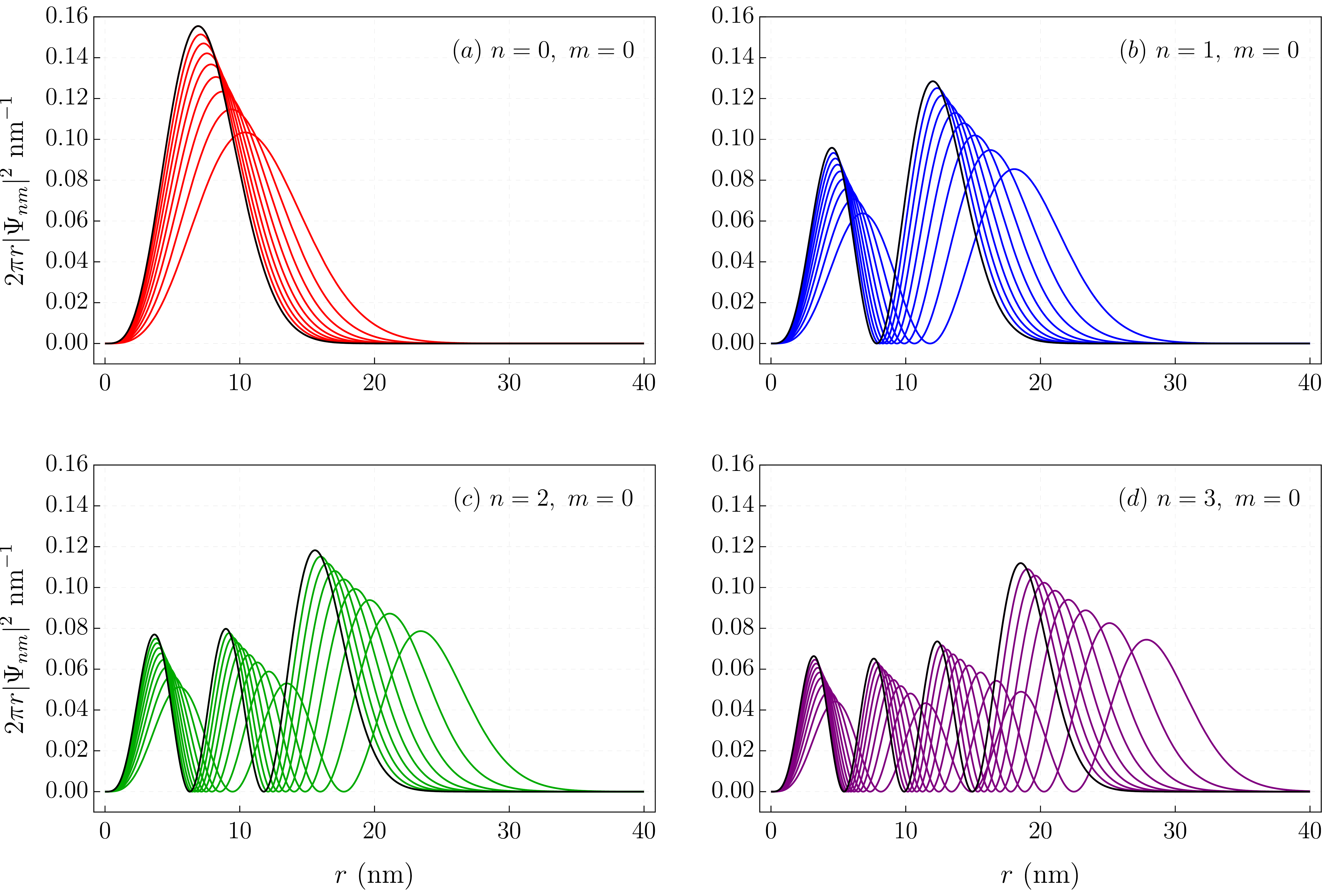}
\caption{Probability distribution function for positive rotations. The parameter values used are $\hbar \omega_0=25.6~\mathrm{meV}$, $r_0=10~\mathrm{nm}$, $\phi=0.5\,h/e$, $B=15~\mathrm{T}$ and $\Omega=0,10,20,\dots,80~\mathrm{THz}$. In (a), we plot the state $\Psi_{00}$ and, in (b), the state $\Psi_{10}$. The black curve in all the graphs represents the value of $\Omega=80$ THz.}
\label{fig:dpr}
\end{figure}
\begin{figure}[tbh]
\centering
\includegraphics[scale=0.15]{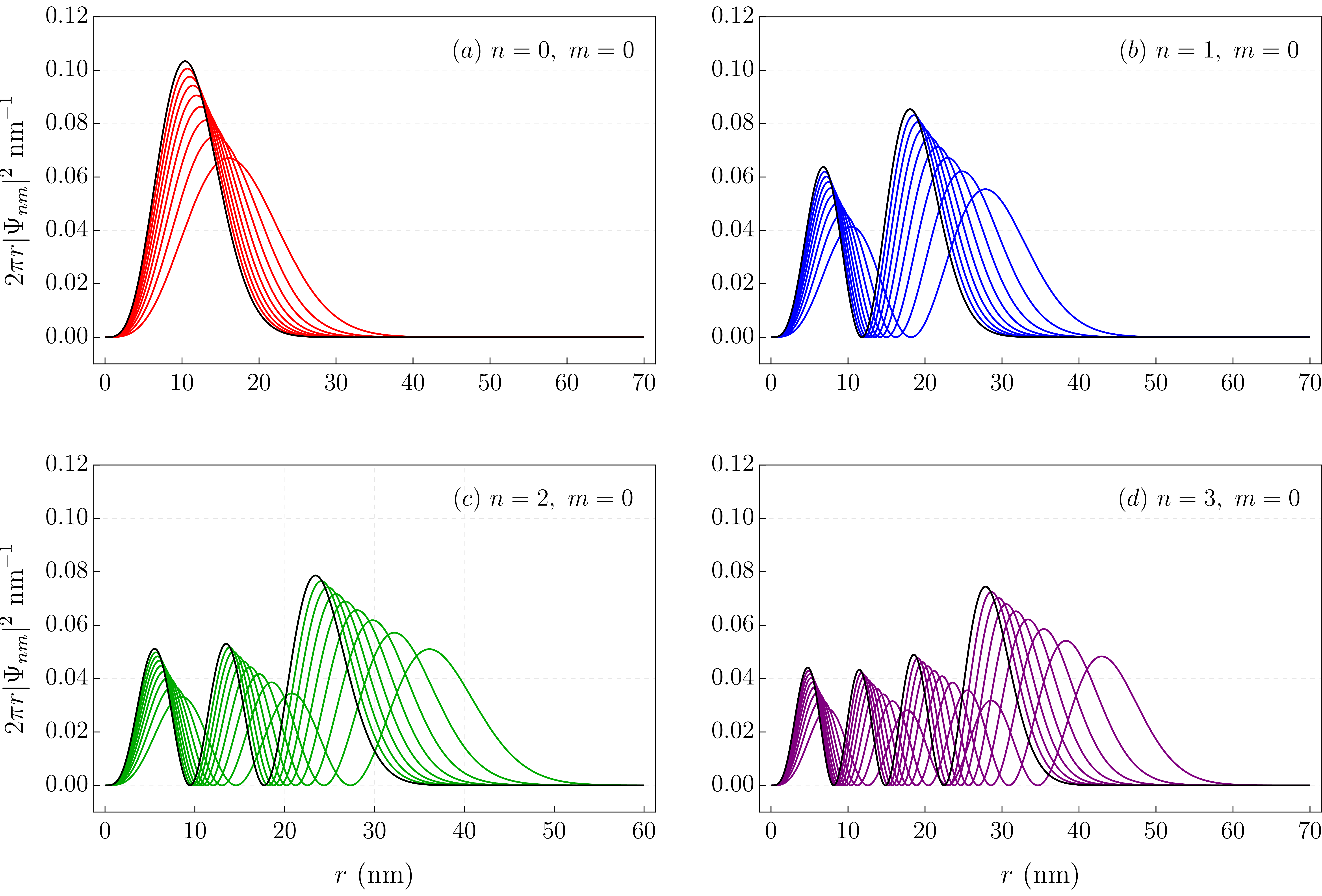}	
\caption{Probability distribution function for negative rotations. The parameter values used are $\hbar \omega_0=25.6\,\mathrm{meV}$, $r_0=10~\mathrm{nm}$, $\phi=0.5 \,(h/e)$, $B=15\,\mathrm{T}$ and $\Omega= -16, -14, -12, \dots, 0~\mathrm{THz}$. In (a), we plot for the state $\Psi_{00}$ and, in (b), for the state $\Psi_{10}$. The black curve in all the graphs represents the value of $\Omega=0$.}
\label{fig:fdn}
\end{figure}
\begin{figure}[tbh]
\centering
\includegraphics[scale=0.50]{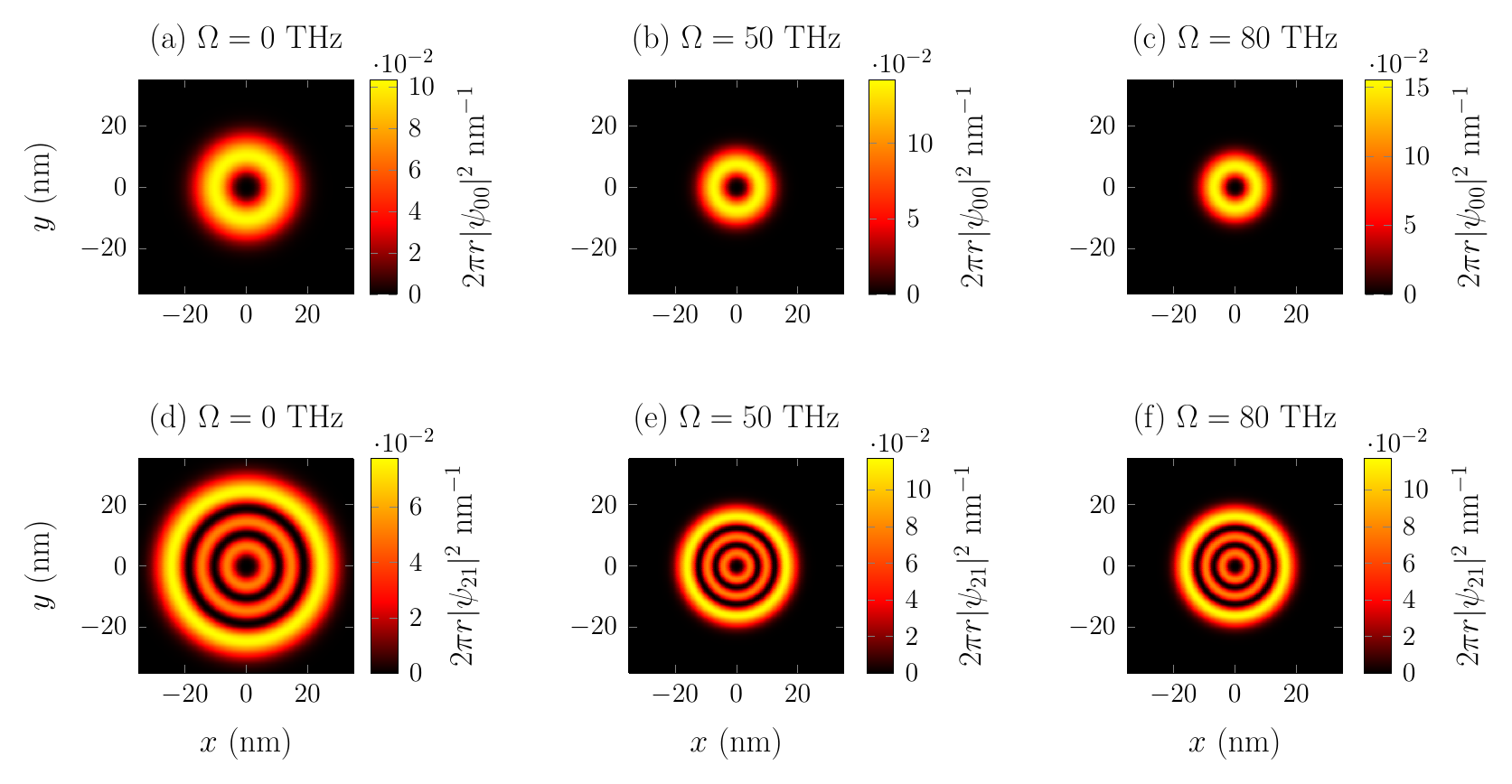}
\caption{Probability distribution function for different rotations. The parameter values used are $\hbar \omega_0=25.6\,\mathrm{meV}$, $r_0=10\,\mathrm{nm}$, $\phi=0.5\,(h/e)$, and $B=15\,\mathrm{T}$. Figs. (a), (b), and (c) are plotted for the state $\Psi_{00}$ and Figs. (d), (e), and (f) for the state $\Psi_{21}$.}
\label{fig:densityrotpositve}
\end{figure}
\begin{figure}[tbh]
\centering
\includegraphics[scale=0.50]{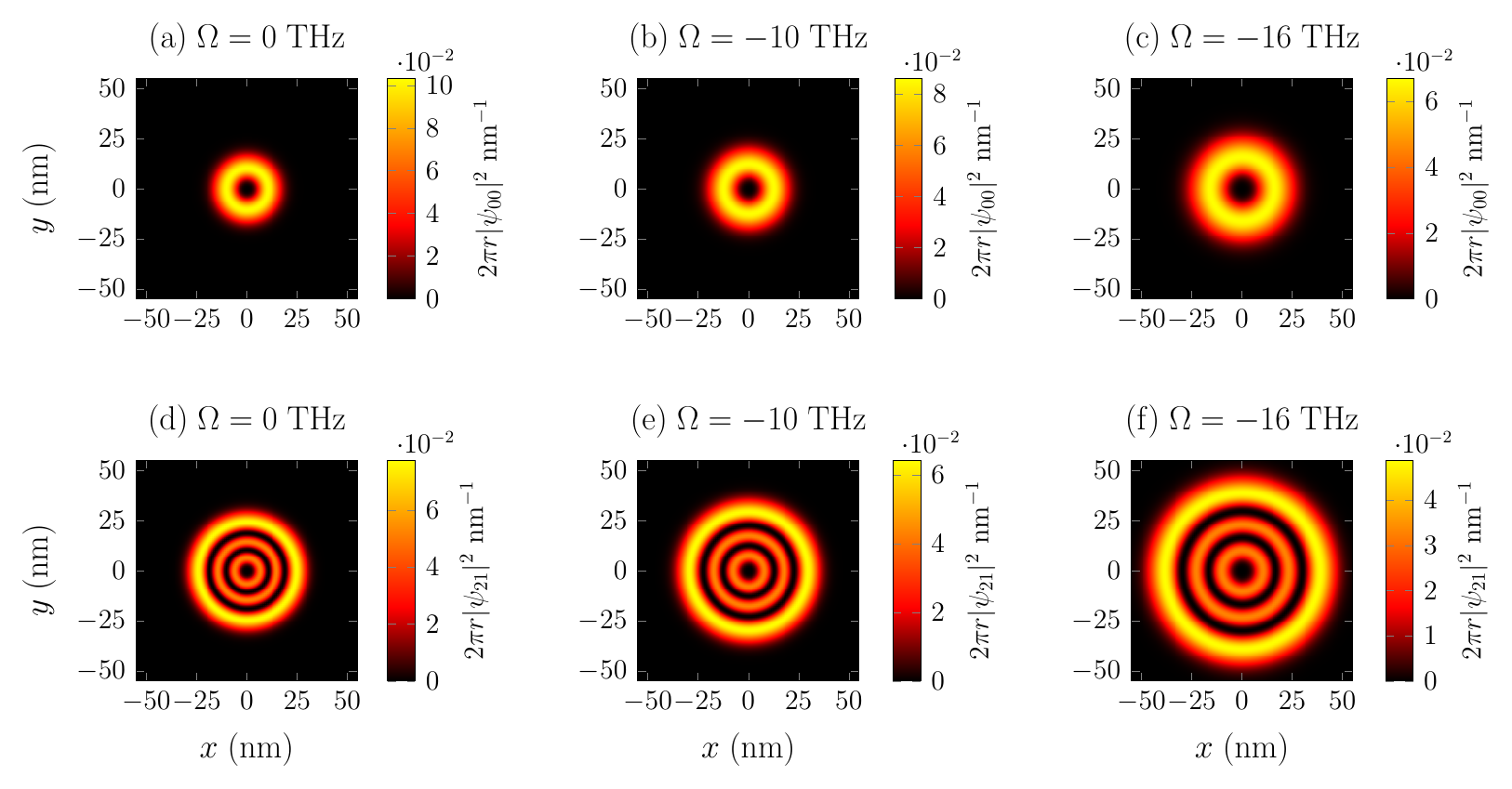}
\caption{Probability distribution function for negative rotations. The parameter values used are $\hbar \omega_0=25.6\,\mathrm{meV}$, $r_0=10\,\mathrm{nm}$, $\phi=0.5\,\mathrm{h/e}$, and $B=15\,\mathrm{T}$. Figs. (a), (b), and (c) are plotted for the state $\Psi_{00}$ and Figs. (d), (e), and (f) for the state $\Psi_{21}$.} 
 \label{fig:densityrotnegative}
\end{figure}
\begin{figure}[tbh]
\centering
\includegraphics[scale=0.9]{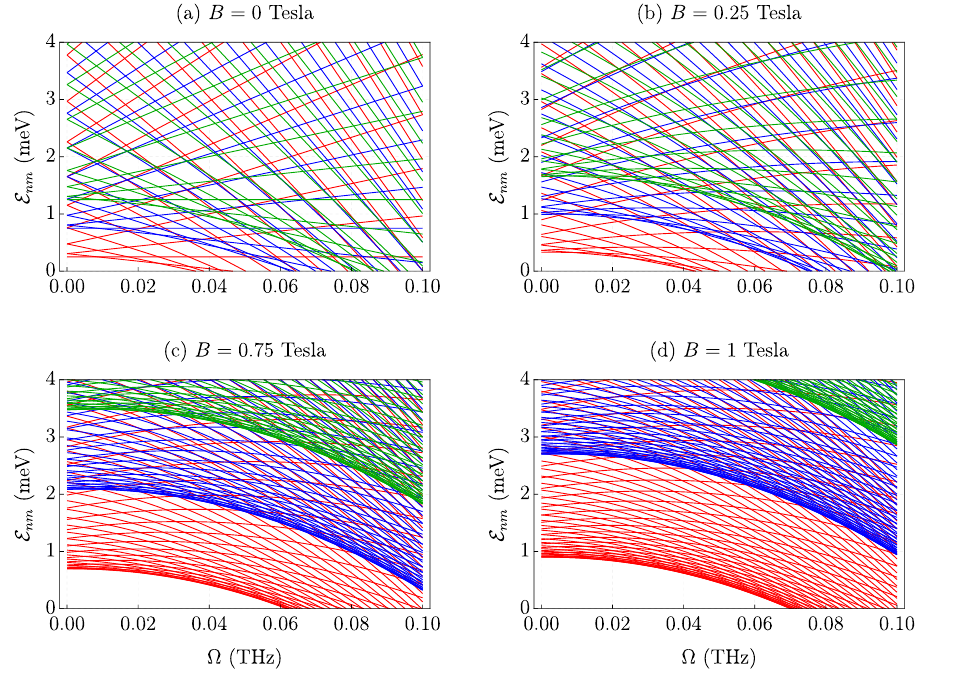}
\caption{Energy graph as a function of angular velocity for the three lowest sub-bands. The parameter values used are $\hbar\omega_{0}=0.50$\,meV, $\phi=0\,(h/e)$, $r_{0}=1000$~nm, and with $m$ ranging from $-300$ to $1300$ with an interval of $10$. The red, blue, and green colors are for the quantum numbers $n=0$, $n=1$, and $n=2$, respectively.}
\label{fig:EnergiaOmega}
\end{figure}
\begin{figure}[tbh]
\centering
\includegraphics[scale=0.9]{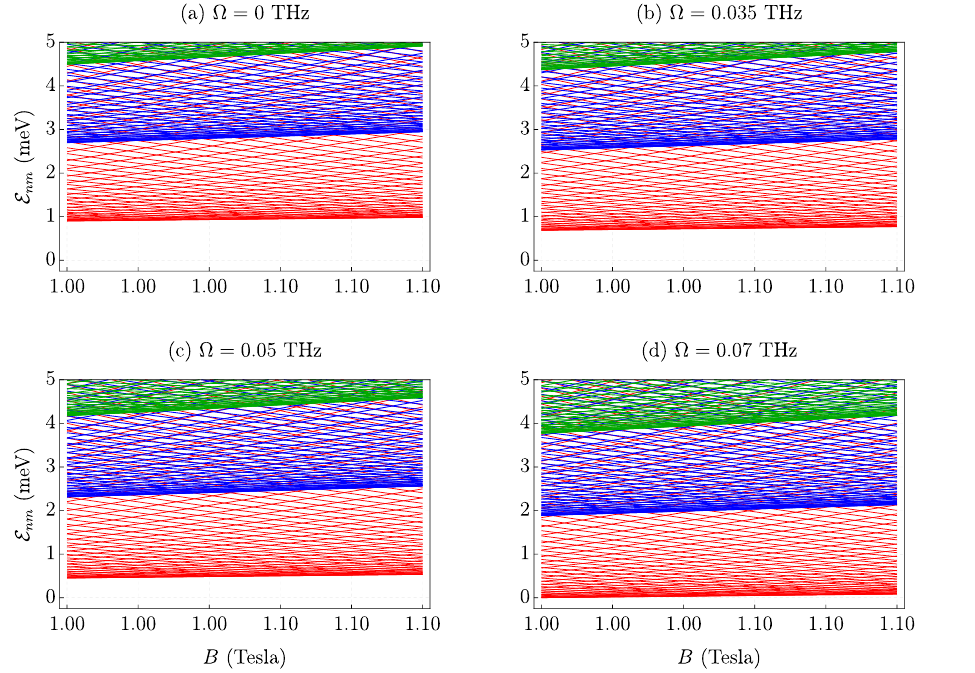}
\caption{Energy as a function of magnetic field for the three lowest sub-bands. The parameter values used are $\hbar\omega_{0}=0.50$\,meV, $\phi=0\,(h/e)$, $r_{0}=1000$~nm, and with $m$ ranging from $-300$ to $1300$ with an interval of $10$. The red, blue, and green colors are for the quantum numbers $n=0$, $n=1$, and $n=2$, respectively.}
\label{fig:EnergiaCampoB}
\end{figure}
This result shows that the quantized energy eigenvalue depends on the quantum numbers $n$ and $m$, where $n$ is the quantum number characterizing the particle's motion in the radial direction and $m$ is the angular momentum. Additionally, three frequencies characterize the energy levels. The first one, $\varpi$, depends on the magnetic field, the coefficient $a_{2}$ of the confinement potential, and the angular velocity $\Omega$. The second frequency $\omega^{*}$ depends on the cyclotron frequency $\omega_{c}$ and the rotation angular frequency $\Omega$. Finally, the third frequency $\omega_{0}$ is a characteristic of the confinement potential and depends exclusively on the parameter $a_{2}$ \cite{DANTAS201511}.

The non-inertial effects are present as the wave function depends on the rotation parameter. One way to explore such effects is to sketch some graphs of the probability density $2\pi r|\Psi_{nm}|^{2}$, exploring some angular velocity values, including zero. Comparing positive and negative rotations implies considering the orientation of the angular vector $\Omega$ relative to the $z$ axis, following the right-hand rule. Such analysis determines whether the system is undergoing clockwise or counterclockwise rotational motion. The values of angular velocities will be in terahertz (THz) as a function of the radius $r$ in nanometers, both positive and negative \cite{AoP.2023.459.169547}.

Figure \ref{fig:dpr} displays the probability distribution function of cohesively finding a particle in the 2D quantum ring. It can be observed that for the ground electronic state ($n=0$, $m=0$), as the rotation increases, there is a greater concentration of electronic states near the center of the ring. In Figure \ref{fig:dpr}(b), the rotation effects can also be observed for other states besides the ground state, excited states, in this case, for $n=1$ and $m=0$.  

For negative rotations, the probability density decreases as the angular velocity moves away from zero, and the resonant peak moves away from the center of the ring as shown in Figure \ref{fig:fdn}, a behavior similar to the graph in Figure \ref{fig:dpr}. In this setup and under the given parameters, as the angular velocity increases, the electronic states tend to concentrate towards the center of the ring. Conversely, as the angular velocity decreases, the probability density moves further away from the center, especially for negative $\Omega$ values. 

It's worth noting that the angular velocities analyzed in this section are on the order of terahertz, indicating that high speeds are required to explicitly observe rotating effects in graphical analysis. This was also verified in Refs. \cite{AdP.2022.535.2200371,AoP.2023.459.169547,DANTAS201511}. 
However, even if we do not have a definitive 
answer for this subject; these effects should be verified even at a much lower experimentally meaningful range for the rotating parameter \cite{MERLIN1993421}.

We also elaborated the representation of the normalized probability density in the $xy$ plane, where each color corresponds to a numerical value associated with the density. Figure \ref{fig:densityrotpositve} illustrates a series of normalized probability density distributions in the $xy$ plane considering $r=\sqrt{x^{2}+y^{2}}$. As rotation increases, it is possible to observe that the probability of finding the particle approaches the center ($x=0$ and $y=0$), evidenced by the lighter colors, highlighting the contribution of the centrifugal term.

Figure \ref{fig:densityrotpositve} shows the density function for the states $\Psi_{00}$ and $\Psi_{21}$ as a function of angular velocity for rotation values of $0\,\mathrm{THz}$, $50\,\mathrm{THz}$, and $80\,\mathrm{THz}$, the same quantum state can change the profile of the probability density by varying the angular velocity parameter.

For negative rotations, we have the opposite effect. The ring widens as the angular velocity decreases, so the inner edge moves away from the center ($x=0$ and $y=0$). To make this more evident, we plot in Figure \ref{fig:densityrotnegative} two distinct states $\Psi_{00}$ and $\Psi_{21}$ for different negative values of angular velocities equal to $0\,\mathrm{THz}$, $-10\,\mathrm{THz}$, and $-16\,\mathrm{THz}$. The yellowish region with a ring geometry in both the graph in Figures \ref{fig:densityrotpositve} and \ref{fig:densityrotnegative} is where the particle is most likely to be found. Also note that as the angular velocity varies, this region is affected concerning the center of the quantum ring.

Likewise, the energy spectrum also undergoes rotation effects. Let's rewrite Equation (\ref{eq:energirrnm}) more explicitly as
\begin{align}
	\mathcal{E}_{nm}  &=\left(n+\frac{1}{2}\sqrt{\left(m-l\right)^{2}+\frac{2\mu a_{1}}{\hbar^{2}}}+\dfrac{1}{2}\right) \hbar\sqrt{\omega_{c}^{2}+4\Omega\omega_{c}+\omega_{0}^{2}}
	\notag\\& -\frac{1}{2}\left(m-l\right)\hbar\left(\omega_{c}+2\Omega\right)-\dfrac{\mu}{4}\omega_{0}^{2}r_{0}^{2}\label{eq:Enmrotate}
\end{align}
to then investigate the effects due to rotation. In the particular case where rotation is zero, we restore the energies (\ref{eq:Enm}) from Sec. \ref{sec2}.
On the other hand, if we assume $\omega_{c}=0$ in Equation (\ref{eq:Enmrotate}), we obtain the energy spectrum
\begin{align}
	\mathcal{E}_{nm}^{\prime} =\left(n+\dfrac{1}{2}\sqrt{\left(m-l\right)^{2}+\frac{2\mu a_{1}}{\hbar^{2}}}+\dfrac{1}{2}\right)\hbar\omega_{0}-\left(m-l\right)\hbar\Omega-\dfrac{\mu}{4}\omega_{0}^{2}r_{0}^{2},
\end{align}
showing that rotation introduces a shift in the particle's energy levels. 

Let's study the energies (\ref{eq:Enmrotate}) and investigate the physical implications of rotation and other physical parameters. Figure \ref{fig:EnergiaOmega} shows the behavior of the energy eigenvalue in units of $\mathrm{meV}$ from Equation (\ref{eq:Enmrotate}) for the three lowest sub-bands ($n=0,1,2$) as a function of angular velocity in $\mathrm{THz}$.
Curves of the same color represent a sub-band indicated by the quantum number $n$, each curve represents a different state of $\Psi_{nm}$ for distinct $m$. The states can occupy energy levels from another sub-band, which is more evident in the first graph for the null magnetic field value, $B=0$. We also observe a greater concentration of electronic states in the lower part of each sub-band, i.e., the degeneracy is higher; as the energy increases, this degeneracy decreases. For higher values of $\Omega$, we observe that the distribution of states predominantly concentrates on lower energy levels. Figure \ref{fig:EnergiaOmega} was constructed for the following fixed magnetic field values: $B=0~\mathrm{Tesla}$, $B=0.25~\mathrm{Tesla}$, $B=0.75~\mathrm{Tesla}$, and $B=1~\mathrm{Tesla}$, as the aim is to analyze the effects that angular velocity has on the energy eigenvalues. Note that for a given chosen magnetic field value, the numerical value of the energy minima of each subband decreases as $\Omega$ increases.

In parallel, we can see an opposite effect of energy levels regarding the magnetic field, represented by the predominant occupation of these concentrations of states at higher energy levels in the face of small variations in the magnetic field values. As illustrated in Figure \ref{fig:EnergiaCampoB}, it is observed that the lower part of the sub-bands, where there is a higher concentration of electronic states, shifts to higher energy values as $B$ increases, while it shifts to lower energy values as $\Omega$ increases. 
Each graph in Figure \ref{fig:EnergiaCampoB} was constructed assuming a fixed rotation parameter since, in this case, we are interested in studying how electronic states evolve as a function of the magnetic field. These two configurations show that one of the effects of rotation is the removal of degeneracy in the inner part of each sub-band \cite{Pereira_2023}.

Finally, we want to emphasize that the effects of rotation on the energy levels and wave functions of electrons in 2D QRs shown in this section lead to various effects on physical properties, as mentioned throughout this paper. Specifically, these effects are significant in the photoionization process. The graphs illustrating the effect of rotation on the PCS of 2D QRs can be found in our referenced work with a co-author \cite{CTP.2024.76.105701}.

\section{Conclusions}
\label{sec5}

In this work, we have examined various aspects of electronics in low-dimensional semiconductor systems, specifically focusing on quantum rings both in the absence of rotation and under rotating effects. We began by reviewing the theoretical studies conducted by Tan and Inkson, which explored the confinement potential in the presence of a magnetic field and a flux tube passing through the center of the ring. We have outlined the main properties of this system, defining the magnetic field configuration and specific sizes of the confining radial potential, writing the system's Hamiltonian, and subsequently deriving the radial equation of motion. We solved this equation to find the normalized energies and wave functions. These results were discussed through graphs illustrating the energy levels as a function of the quantum number $ m $, the magnetic field $ B $, the AB flux $ \phi/\phi_{0} $, and angular velocity $ \Omega $. 

Additionally, we investigated the optical properties of the quantum system, particularly the photoionization cross-section, which is an important optical property for studying low-dimensional semiconductor structures. The results regarding the incident photon energy and the AB flux for the two lowest optical transitions reveal different values when certain ring parameters are varied, as the energy eigenvalues of the states participating in the optical transitions change. Our findings show that the photoionization cross-section in QRs strongly depends on parameters such as the average radius and the AB flux. Furthermore, we observed that the first transition $(n=0, m=0)$ to $(n=0, m=-1)$ has a higher probability of leading to the photoionization process compared to the second transition $(n=0, m=0)$ to $(n=0, m=1)$.

Subsequently, we studied the 2D quantum ring under rotation effects. We found that rotation is responsible for the displacement of the probability density of finding the particle in the quantum ring, showing different results for the orientation of the angular velocity vector according to the right-hand rule. The angular velocity variation showed interesting results in the energy eigenvalues, both for positive and negative angular velocities. This, of course, shows that the rotating effect in the Tan-Inkson model may cause important changes in the form and size of the amplitudes and signs of various physical properties, such as the refractive index changes, absorption coefficients, photoionization cross-section, magnetization, and persistent currents, and thermodynamics properties. As mentioned in this review, some of these studies have already been completed, while others have yet to be completed.

The study of QRs and other types of mesoscopic structures is of great importance for various reasons, including the technological evolution of components at the nanometer scale. This rise is a key factor driving their investigation and application in devices with nanostructured configurations. We believe our study opens new perspectives for future investigations in this field, covering electronic, optical, and thermodynamic properties. Additionally, it provides motivation to investigate novel trends in the photoionization process of QRs, such as the presence of impurities \cite{pal2019impurity,PBB.2008.77.045317,peter2021magneto}, spin-orbit interactions \cite{AOP.2013.339370,Smolkina_2019}, and the effects arising from rotation, which are discussed in detail here.

Finally, we would like to emphasize that this work is more than just a simple review. We have approached and presented the results from novel perspectives, and by comparing them with existing literature, we aim to provide the reader with a deeper understanding of the subject. In fact, some of our graphs and numerical results are genuinely novel.

\section*{\label{sec:acknw}Acknowledgement}

This work was partially supported by the Brazilian agencies CAPES, CNPq, and FAPEMA. E. O. Silva acknowledges CNPq Grant 306308/2022-3, FAPEMA Grants APP-12256/22 and UNIVERSAL-06395/22. F. S. Azevedo acknowledges
CNPq Grant No. 153635/2024-0.  This study was financed in part by the Coordena\c{c}\~{a}o de Aperfei\c{c}oamento de Pessoal de N\'{\i}vel Superior - Brazil (CAPES) - Finance Code 001.

\bibliographystyle{apsrev4-2}
\input{Article.bbl}

\end{document}

%% file: Article.bbl
%